\documentclass[preprint,3p,twocolumn]{elsarticle}
\usepackage{geometry}
\usepackage{fancyhdr}

\biboptions{numbers,sort&compress}
\usepackage{amsmath}
\usepackage{amssymb}
\usepackage{graphicx}
 \usepackage{gensymb}
\usepackage{pdfpages}

\begin{document}

\begin{frontmatter}

\title{Grain growth competition during melt pool solidification -- \\ Comparing phase-field and cellular automaton models}

\author[indus,imdea]{S.M.~Elahi}
\author[imdea]{R.~Tavakoli}
\author[indus,imdea]{I.~Romero}
\author[imdea]{D.~Tourret$^*$}

\address[indus]{Universidad Polit\'ecnica de Madrid, E.T.S. de Ingenieros Industriales, Madrid, Spain}
\address[imdea]{IMDEA Materials, Madrid, Spain}

 \cortext[corr]{Corresponding author; Email address: damien.tourret@imdea.org}

\begin{abstract}

A broad range of computational models have been proposed to predict microstructure development during solidification processing but they have seldom been compared to each other on a quantitative and systematic basis. In this paper, we compare phase-field (PF) and cellular automaton (CA) simulations of polycrystalline growth in a two-dimensional melt pool under conditions relevant to additive manufacturing (powder-bed fusion). We compare the resulting grain structures using local (point-by-point) measurements, as well as averaged grain orientation distributions over several simulations. We explore the effect of the CA spatial discretization level and that of the melt pool aspect ratio upon the selected grain texture. Our simulations show that detailed microscopic features related to transient growth conditions and solid-liquid interface stability (e.g. the initial planar growth stage prior to its cellular/dendritic destabilization, or the early elimination of unfavorably oriented grains due to neighbor grain sidebranching) can only be captured by PF simulations. The resulting disagreement between PF and CA predictions can only be addressed partially by a refinement of the CA grid. However, overall grain distributions averaged over the entire melt pools of several simulations seem to lead to a notably better agreement between PF and CA, with some variability with the melt pool shape and CA grid. While further research remains required, in particular to identify the appropriate selection of CA spatial discretization and its link to characteristic microstructural length scales, this research provides a useful step forward in this direction by comparing both methods quantitatively at process-relevant length and time scales.

\end{abstract}


%%%%%%%%%%%%%%%%%%%%%%%
\end{frontmatter}
\thispagestyle{fancy} 

%=================================================================

\section{Introduction}
\label{sec:intro}

The formation of grain structures during the solidification of metals and alloys has a crucial effect on the final material properties and the mechanical integrity of structural components. 
In additive manufacturing (AM), just like in conventional manufacturing processes, grain textures lead to strongly anisotropic material properties \cite{carroll2015anisotropic, kirka2017mechanical, gordon2018fatigue, mooney2019process, hosseini2019review}.
The flexibility afforded by AM to locally modulate processing conditions (e.g. through heat source power or velocity) provides a strong motivation for trying to control local microstructure and local mechanical properties of AM components.
However, in order to unleash this disruptive potential in fusion-based AM, important gaps remain in our understanding of grain texture selection during solidification in a melt pool.

Grain textures first emerge during grain nucleation and growth competition during the solidification stage.
Grain growth competition involves a complex interaction between thermo-solutal conditions and intrinsic material properties.
On the one hand, thermal gradients and growth velocities determine the type of microstructure (e.g. planar, cellular, dendritic).
On the other hand, the intrinsic crystalline structure dictates solid-liquid interface anisotropy and preferred growth directions. 
Early theories considered that grains with their preferred growth direction ($\langle100\rangle$ in most cubic crystals) best aligned with the main temperature gradient direction had to be favored in the growth competition against neighboring grains with higher misorientation to the temperature gradient \cite{walton1959}.
While this is, on average, considered to be true, later experimental observations highlighted more complicated growth competition mechanisms that did not follow this simple picture \cite{dsouza2002morphological,wagner2004grain,zhou2008mechanism}.
Recent computational studies \cite{li2012phase, tourret2015growth, takaki2016two, tourret2017grain} have provided major advances into the fundamental understanding of grain growth competition mechanisms, focusing on idealized temperature conditions, namely with a one-dimensional temperature gradient.

Deviations from such an idealized thermal field may have important consequences on selected microstructures. 
For instance, in directional (Bridgman) solidification, a small curvature of the isotherms due to lateral heat extraction \cite{mota2015initial} or a global shift of the isotherms due to thermal inertia of the experimental setup \cite{song2018thermal}, may lead to strong differences in microstructures, e.g., leading to the lateral drift of cellular/dendritic patterns and heterogeneities in the selected primary spacings \cite{song2018thermal, mota2021effect}, which strongly affects the dynamics of grain growth competition \cite{mota2021effect}.
Yet, these deviations from a one-dimensional temperature field are often neglected in fundamental studies of polycrystal growth, which often use idealized temperature fields, e.g. the classical frozen temperature approximation \cite{li2012phase, tourret2015growth, takaki2016two, tourret2017grain, mota2021effect}.
One objective of the present study is to perform simulations of polycrystalline solidification within a multi-dimensional thermal field relevant to AM, in order to analyze the mechanisms of grain growth competition in a melt pool.

While experimental studies of carefully controlled bi-crystal growth competition are possible \cite{liu2013dependence,yu2014anomalous, hu2018competitive, wang2019competitive}, they remain challenging to perform.
Meanwhile, quantitative simulations of solidification have emerged that allow thorough studies of microstructure selection while exploring individual parameters -- e.g. crystal orientations, temperature gradients, and cooling rates -- independently.
A host of models for solidification have been developed using a broad palette of numerical methods, including front tracking \cite{zhao2001front, li2003fixed}, Monte Carlo \cite{rodgers2017simulation, rodgers2021simulation}, phase-field-crystal \cite{jreidini2021orientation}, level set \cite{kim2000computation, gibou2003level}, grain envelope \cite{steinbach1999three, souhar2016three, viardin2017mesoscopic}, needle network \cite{tourret2016three, isensee2022convective}, cellular automaton \cite{rappaz1993probabilistic, gandin1994coupled}, and phase-field \cite{boettinger2002phase, tourret2022phase} approaches. 
In this study, we focus on cellular automaton and phase-field methods, which are by far the most utilized approaches for simulating grain structure predictions in solidification, to a great extent due to their generality and simplicity.

At the scale of individual dendrites or dendritic arrays, phase-field (PF) is an outstanding method for simulating the evolution of morphologically complex interfaces based only on thermodynamics and kinetics considerations \cite{boettinger2002phase, steinbach2009phase, tourret2022phase}. 
The PF method has proven especially successful in the field of solidification, where ``mesoscale'' formulations have been developed that remain faithful to the sharp-interface problem even for diffuse interfaces much wider than the actual interface width \cite{karma1998quantitative, echebarria2004quantitative}. 
However, the spatial discretization required for well-converged quantitative simulations is intrinsically limited by the microstructural length scale, i.e. the local interface curvature \cite{echebarria2004quantitative, steinbach2009phase, shibuta2015solidification}.
These limitations can, to some extent, be alleviated using advanced implementations, e.g. using massive parallelization on graphics processing units (GPUs) \cite{shimokawabe2011peta}, but resulting simulations remain computationally expensive.
We have recently shown that well-converged PF simulations of polycrystalline solidification in AM-relevant conditions could be achieved at the scale of the entire melt pool in two dimensions (2D) \cite{elahi2022multiscale}.
Yet, even though the resulting simulation was performed using reasonable computing resources (one node equipped with eight Nvidia GPUs), such simulations remain quite challenging computationally (e.g. involving over one billion grid points).

At the larger scale of entire grain structures or even entire components, cellular automaton (CA) models provide a good compromise between efficiency and physics-based considerations \cite{rappaz1993probabilistic, gandin1994coupled, chen2016three, rai2016coupled, koepf2019numerical, lian2019cellular, mohebbi2020implementation, teferra2021optimizing}.
In CA models, grains are constructed from polyhedral building blocks, whose vertices mark the preferred crystallographic growth directions. 
Growth velocities in these directions follow simplified, yet physics-based, kinetic laws for crystal growth -- e.g. using Ivantsov-based relations or power laws relating local supersaturation or undercooling \cite{ivantsov1947temperature, kurz1986theory}. 
As such, CA models include more adjustable phenomenological parameters than PF models. 
The significant computational advantage -- and hence upscaling potential -- of CA over PF resides in their relatively much coarser spatial discretization.
Advanced schemes have been developed to couple CA with macroscopic thermomechanical solvers, e.g. using finite elements (FE) \cite{gandin1994coupled, carozzani20113d, chen2016three, koepf2019numerical, teferra2021optimizing}, thus enabling the combination of solidification thermodynamics and kinetics (CA) with macroscopic transport of heat, solute and mechanics (FE).
Notably, CA with finer grid sizes have also been used to simulate the growth of individual dendrites \cite{wang2003model, yin2011simulation, choudhury2012comparison, eshraghi2015large}. 
However, additional terms are required to capture the effect of local interface curvature, and the cell size needed to appropriately represent dendrite tips somewhat reduces their computational advantage.

A key motivation for the development of CA models for solidification in the 1990's was specifically the prediction of grain textures via grain growth competition \cite{gandin1995grain, rappaz1996prediction, takatani2000ebsd, carter2000process}.
Early articles suggested that the appropriate length scale for the typical CA cell size ($h_{\rm CA}$) was the secondary dendrite arm spacing \cite{gandin19973d}.
However, until recently \cite{pineau2018growth,dorari2022growth}, the effect of $h_{\rm CA}$ on the predicted grain growth competition, and its link to microstructural length scales, was not carefully studied.
As a result, the grid size has often been selected as an adjustable parameter based principally on numerical convenience.
Recent studies \cite{pineau2018growth,dorari2022growth} have shown that CA simulations lead to results closest to PF when $h_{\rm CA}$ is of the same order as the ``height'' (undercooling) difference between the two competing grains within the temperature gradient \cite{pineau2018growth} or comparable to the active secondary dendritic spacing within competing grains \cite{dorari2022growth}.
These studies remain to be extended to polycrystalline microstructures in multi-dimensional temperature fields.
A key objective of the present study is the quantitative comparison of PF and CA predictions of grain structures in AM-relevant conditions, performed over a statistically significant number of grains, in order to assess the effect of $h_{\rm CA}$ and provide further guidelines for its selection.

Hence, in this paper, we compare directly and quantitatively the grain structure predicted by PF and CA models for similar thermal conditions relevant to AM. 
We perform these calculations in two dimensions, but at the entire melt pool scale, and repeat most of them several times with different initial grain structures, in order to obtain a statistically meaningful picture of grain growth competition in a two-dimensional temperature field. 
In particular, these results allow us to discuss (1) the mechanisms of unsteady dendritic grain growth competition in an AM-relevant 2D thermal field, (2) the advantages, but also the limitations of upscaling PF simulations using CA, and (3) the  appropriate choice of cell size for a reasonable matching between CA and PF predictions.

%=================================================================

\section{Methods}

With the objective to compare PF and CA simulations of melt pool solidification, we consider material and laser properties from a previous study on powder-bed fusion of Inconel 718 alloy \cite{elahi2022multiscale}.
Therein, temperature-dependent alloy properties --- namely heat capacity, density, thermal conductivity --- are calculated using the CalPhaD method (ThermoCalc, TCNI8 database) or tabulated from experimental literature. 
The steady-state thermal field in the melt pool region is calculated using finite elements (FE), as described with detail in Ref.~\cite{elahi2022multiscale}.
We use the resulting three-dimensional (3D) thermal field to extract two-dimensional sections, namely longitudinal and transversal, which we use as the temperature field in either PF or CA simulations.
In the subsequent simulations of solidification, we use alloy properties and phase diagram for a binary Ni-5wt\%Nb alloy given in Ref.~\cite{elahi2022multiscale}.
Importantly, we consider that solidification occurs solely by epitaxial growth from the bottom of the melt pool, and neglect nucleation, thus focusing on the growth competition and avoiding the addition of further adjustable parameters (e.g. nuclei density and nucleation undercooling).

The restriction to 2D simulations of polycrystalline solidification is arguably a limitation. However, previous studies have shown that grain growth competition mechanisms are appropriately reproduced by 2D simulations when compared to quasi-2D three-dimensional simulations \cite{tourret2017grain}.
Hence, even though the full space of 3D orientations cannot be scanned in these 2D simulations, we still expect that they should be able to capture -- even if only qualitatively -- phenomena such as the grain texture transition as a function of the melt pool aspect ratio \cite{jadhav2019influence,higashi2020selective}.
Conceptually, both PF and CA models used here are readily applicable in three-dimensions \cite{tourret2017grain, chen2016three}. However, the fine spatial discretization required for PF simulations (here of order 5 to 10 nm) prevents the execution of dendrite-resolved quantitative simulations in 3D at the scale of the entire melt pool, even using massive computational resources (already requiring over a billion grid points in the current longitudinal section simulation \cite{elahi2022multiscale}). From the CA perspective, full melt pool and even multi-pass simulations are achievable \cite{chen2016three}. However, except for bicrystalline 2D simulations in a one-dimensional temperature field \cite{pineau2018growth, elahi2022multiscale}, the accuracy of CA simulations of polycrystalline grain growth competition in AM-relevant conditions has not been addressed to date – which is a key objective of the current article.

%=========================================

\subsection{Thermal field}
\label{sec:method:thermal}

Based on the results of 3D FE, we extract 2D longitudinal (i.e. within a horizontal plane that contains the heat source location and its path) and cross sections (i.e. within a horizontal plane normal to the scanning direction) of the temperature field in the melt pool region.
Then, we approximate the temperature field, $T$, with simple analytical formulae that consider an elliptical shape of liquidus and solidus isotherms.

In terms of grain growth competition during solidification, the most important region, where the temperature field needs to be described accurately enough, is between the solidus ($T_S=1554$\,K) and liquidus ($T_L=1625$\,K) temperatures.
Therefore, the analytical fitted functions are chosen to obtain a reasonable description of the time-dependent location of $(T=T_L)$ and $(T=T_S)$, and hence of the temperature gradient in the freezing range, regardless of the approximation accuracy below $T_S$ or above $T_L$.

\subsubsection{Longitudinal section}

The analytical approximation of the thermal field longitudinal section is similar to that used and discussed in Ref.~\cite{elahi2022multiscale}. Specifically, the temperature field is defined in polar
coordinates through the function 

\begin{align}
\label{eq:Tlong}
T(r, \theta) & = T_L\, + \, (T_0 - T_L)\,
\frac{r-r_L(\theta)}{r_S(\theta)-r_L(\theta)} ,
\end{align}\\
where $r_S(\theta)$ and $r_L(\theta)$ are respectively the radii of the location of solidus and liquidus isotherms at a given angle $\theta = \tan^{-1}|(y-y_0)/(x-x_0)|$, where $(x_0,y_0)$ is the location of the center of the ellipses, and $r=\sqrt{(x-x_0)^2+(y-y_0)^2}$.
The solidus and liquidus isotherms are modeled as two ellipses with

\begin{align}
r_L(\theta) & = \sqrt{\frac{(l_L\; d_L)^2} { \big(d_L \cos
(\theta)\big)^2 + \big( l_L\sin (\theta)\big)^2}}
\label{eq:rl},\\ %
r_S(\theta) & = \sqrt{\frac{(l_S\; d_S)^2} { \big(d_S \cos
(\theta)\big)^2 + \big( l_S\sin (\theta)\big)^2}} ,
\label{eq:rs}
\end{align}
with ($l_L$, $l_S$) the lengths and ($d_L$, $d_S$) the depths of solidus (subscript $S$) and liquidus ($L$) isotherms, respectively.
The resulting temperature field $T$ is shifted in the direction~$x$ with the velocity of the laser beam $V_{\rm L}=0.1~$m/s.

\subsubsection{Cross-section}
\label{sec:temp_cs}

The cross-section temperature is also approximated using elliptical solidus and liquidus isotherms, which shrink non-linearly with time, in order to provide a reasonable fit to the FE results.
In this case, the temperature is interpolated linearly between the temperature at the center of the melt pool $T_0$ and the solidus temperature $T_S$ as

\begin{align}
\label{eq:Tcross}
T(r,\theta)&=T_0(\theta,t)+\big(T_S-T_0(\theta,t)\big)\frac{r}{r_S(\theta,t)} ,
\end{align}

\noindent
with the location of the solidus temperature isotherm given by the equation of an ellipsis

\begin{align}
r_S(\theta)&=\sqrt{\frac{(w_S(t)\; d_S(t))^2}{(d_S(t)\cos(\theta))^2+(w_S(t)\sin(\theta))^2}}\ .
\end{align}

The time-dependent half-width $w_S$ and depth $d_S$ of this isotherm are described by

\begin{align}
w_S(t) &= w_0  \left[1- \big(t/t_f\big)^{n_w} \right] ,\\
d_S(t) &= d_0  \left[1- \big(t/t_f\big)^{n_d} \right] , 
\end{align}
where $w_0$ and $d_0$ are the initial (i.e. maximal) half-width and depth of the melt pool, and 
$t_f$ is a time at which both $w_S$ and $d_S$ reach zero.
The time evolution of $w_S$ and $d_S$ between $(w_S,w_L)=(w_0,d_0)$ at $t=0$ and $w_S=w_L=0$ at $t=t_f$ is parametrized using exponents $n_w$ and $n_d$ (linear evolution if exponent equals 1, or late and steep evolution for higher exponents).
Parameter values $w_0=128~\mu$m, $d_0=96~\mu$m, $t_f=0.00255$~s, $n_w=20$, $n_d=3$ are adjusted for a good fit to the FE results (see Supplementary Material -- Section~A -- Fig.~A.2.a).

The time evolution of the temperature at the center of the melt pool $T_0(\theta,t)$ is described in order to provide a reasonable location of the liquidus temperature location by radial interpolation between $T_0(\theta,t)$ at $r=0$ and $T_S$ at $r=r_S$.
For our specific FE results, at the early stage of melt pool shrinking, $T_0(\theta,t)$ is well described by a $\theta$-dependent function

\begin{align}
T_1(\theta) = T_w - (T_w-T_d) \sin(\theta), \label{eq:T1}
\end{align}
where $T_w$ and $T_d$ are values of $T_0$ extrapolated radially from the location of $T_S$ and $T_L$ along the width and depth, respectively, and using a simple sine interpolation between $T_w$ and $T_d$ along $\theta$.
At the later stage of the melt pool shrinking, $T_0(\theta,t)$ is well described by a time-dependent function

\begin{align}
T_2(t) = T_L - R\, (t-t_L) ,
\end{align}
where $R$ is a constant cooling rate and $t_L$ is an approximate time at which $T_0$ approaches $T_L$ (see Supplementary Material -- Section~A -- Fig.~A.2.b).
Between early and late stages, $T_0(\theta,t)$ is interpolated in time between $T_1(\theta)$ and $T_2(t)$ using a sigmoid (hyperbolic tangent) function

\begin{align}
T_0(\theta,t) &= \frac{T_1(\theta)+T_2(t)}{2} \nonumber\\
 &- \frac{T_1(\theta)-T_2(t)}{2}\; \tanh\left(\frac{t-\tau(\theta)}{\sigma(\theta)}\right) \label{eq:sigmoid},
\end{align}
in which the time and duration of the sigmoid transition from $T_1$ to $T_2$ are parametrized using 

\begin{align}
\tau(\theta) &= \tau_w - (\tau_w-\tau_d)\sin(\theta) ,  \label{eq:tau}\\
\sigma(\theta) &= \sigma_w - (\sigma_w-\sigma_d)\sin(\theta) \label{eq:sigma} .
\end{align}
Here, $\tau_w$ and $\sigma_w$ correspond to optimal values along the width ($\theta=0$), whereas $\tau_d$ and $\sigma_d$ are optimal values along the depth ($\theta=\pi/2$), interpolated along $\theta$ using a simple sine function.
Parameter values 
$T_w = 2700$~K, $T_d = 2450$~K, $R=1.2\times 10^5$~K/s, $t_L=0.00195$~s, $\tau_w = 0.00115$~s, $\tau_d = 0.0007$~s, $\sigma_w = 0.00036$~s, and $\sigma_d = 0.0005$~s 
were found to provide a good match to our FE results in terms of liquidus ($T_L=1625$~K) and solidus ($T_S=1554$~K) isotherms and their time evolution (see Supplementary Material -- Section~A -- Fig.~A.3).

The temperature field given by Eq.~\eqref{eq:Tcross} is arguably approximate and has a high number of adjustable parameters, which may even lead to singularities if not chosen carefully.
As such, it is not intended to provide a general description of melt pool temperature fields. 
A direct use of a tabulated FE-calculated field using an efficient interpolation scheme could have been a more accurate choice.
However, this analytical approximation still provides a reasonable analytical description of the FE-calculated temperature field (see Supplementary Material -- Section~A), which is more than sufficient for our study of grain growth competition.

\subsubsection{Explored thermal configurations}
\label{sec:method:configs}

We simulate one configuration for the longitudinal section, similar to that presented in Ref.~\cite{elahi2022multiscale} and 15 different cross-section configurations, as summarized in Table~\ref{tab:thermal_configs}. For the latter, we consider three melt pool shapes: (1) one corresponding to the original fit to the FE results (i.e. using equations and parameters listed in Section~\ref{sec:temp_cs}), (2) one with a twice wider and twice shallower melt pool (i.e. with $w_0=256~\mu$m, $d_0=48~\mu$m), and (3) one with a twice deeper and twice narrower melt pool (i.e. with $w_0=64~\mu$m, $d_0=192~\mu$m). 
For each of these three melt pool shapes, we consider five initial configurations with different randomly generated initial grain maps. 

\begin{table}[h!]
\centering
\caption{Thermal configurations for cross-section simulations.\label{tab:thermal_configs}}
\begin{tabular}{lccc}
\hline
 Configuration & Domain Size & $w_0$ & $d_0$ \\
 & ($\mu{\rm m}\times\mu{\rm m}$) & ($\mu{\rm m}$) & ($\mu{\rm m}$) \\
\hline
Reference & $257\times97$ & 128 & 96  \\
Wide \& Shallow & $513\times49$ & 256 & 48 \\
Deep \& Narrow & $129\times193$ & 64 & 192 \\
\hline
\end{tabular}
\end{table}

\subsubsection{Initial grain structures}
\label{sec:method:voronoi}

To build these initial grain maps, a total of 1500 grain centers are generated at random locations using a Poisson disk sampling algorithm. 
A random orientation, namely an integer in the range $[1-90]$, is attributed to each grain center.
This integer grain index is then used as the crystalline orientation, in degrees, of the grain with respect to the computational grid frame of reference.
After the Voronoi tesselation step that propagates the orientation of each center to its corresponding polygonal grain, the grid points with temperature above $T_L$, which belong to the melted region, are reinitialized to the liquid state and their grain orientation is ``erased''. 
Solidification calculations for the sixteen configurations are carried out once with PF simulations (Section~\ref{sec:method:pf}) and several times with different grid element sizes with CA simulations (Section~\ref{sec:method:ca}).
It should be noted that the Voronoi-based initial structures are used for simplicity and generality but not intended to represent a realistic AM microstructure -- e.g. from previous layer building -- but rather used as a test case to compare PF and CA predictions in identical configurations.

\subsection{Phase-Field}
\label{sec:method:pf}

The PF model used here is similar to that described in Ref.~\cite{elahi2022multiscale}.
It is a classical quantitative PF model for dilute binary alloy solidification \cite{echebarria2004quantitative} with nonlinearly preconditioning of the phase field \cite{glasner2001nonlinear, tourret2015growth} considering an externally imposed temperature field  \cite{elahi2022multiscale}.
The solid-liquid interface is assumed to be at equilibrium. 
Neglecting solute trapping restricts its application to a low-to-moderate interface velocity relevant to the lower
velocity range experience in powder-bed fusion processes. 
For the considered laser velocity $V=0.1~$m/s, the interface remains in the appropriate velocity range to neglect any significant solute trapping effects (see discussion in Ref.~\cite{elahi2022multiscale}).
Furthermore, the velocities of isotherms along the cross-section are even lower than those in the longitudinal section \cite{karayagiz2020finite}.
We use a simple polycrystalline extension of the model \cite{elahi2022multiscale} which does not consider any solid-solid motion of the grain boundaries (GB) once formed.
This assumption neglects any further coarsening of the microstructure, e.g. during the intrinsic heat treatment experienced by the microstructure during the deposition of upper layers.
However, here we chose to focus primarily on the epitaxial growth and grain growth competition of columnar grains with different crystal orientations in the melt pool. 

All boundary conditions are set to no-flux on both solute and phase fields. 
The longitudinal section simulation, presented in Ref.~\cite{elahi2022multiscale}, considers a moving frame advancing with respect to the material at the velocity of the laser, hence conserving a temperature field given by Eq.~\eqref{eq:Tlong} fixed within the computational domain.
On the other hand, the cross-section simulations consider a fixed frame with respect to the material and a shrinking melt pool according to Eq.~\eqref{eq:Tcross}.

Phase-field equations are solved with a fully explicit first order in time and second order in space finite difference method on a uniform grid. 
A grid spacing $h_{\rm PF}=5$~nm is used for the spatial discretization for the longitudinal section simulations, following the convergence study performed in Ref.~\cite{elahi2022multiscale}.
In the cross-section simulations, the solid-liquid interface velocities are lower, such that a coarser discretization can be afforded with $h_{\rm PF}=10$~nm.
The time step size is taken to be equal to 0.3 of the stability limit for Laplacian operators. 
With the exception of the grid spacing $h_{\rm PF}$, the initial grain distributions, and the thermal field (Section~\ref{sec:method:thermal}), all parameters are identical to those used in Ref.~\cite{elahi2022multiscale}.

%=========================================
\subsection{Cellular Automaton}
\label{sec:method:ca}

We implemented a standard 2D cellular automaton model to track the envelope of grains \cite{rappaz1993probabilistic, gandin1994coupled, gandin19973d}.
The model uses a uniform grid with a cell size $h_{\rm CA}$.
Each cell is associated with a growing polygon whose vertices are oriented toward the $\langle100\rangle$ ($\langle10\rangle$ in 2D) crystalline directions of the grain and grow at a velocity calculated from the local undercooling.
When the polygon grows enough to capture the center of neighboring cells, the captured cell is activated and its own associated polygon is initialized and starts to grow. 

We considered an eight-neighbor (Moore) capture and a classical decentered square algorithm \cite{gandin19973d, gandin1999three, carozzani20113d, rai2016coupled}.
More advanced algorithms exist that may be more accurate in presence of a high temperature gradient, e.g. considering irregular (kite-shaped) quadrangles accounting for the different local velocities of the individual vertices \cite{takatani2000ebsd, carozzani2012developpement, pineau2018growth, fleurisson2022hybrid}.
However, we chose the decentered square method as it is one of the most widely used due to its simplicity, in particular in recent applications to additive manufacturing \cite{rai2016coupled, lian2019cellular, teferra2021optimizing}, while the approach still reduces grid-induced anisotropy and appropriately captures non-uniform temperature fields \cite{gandin19973d}.

The CA model and algorithm are essentially identical to those of Refs~\cite{gandin19973d,rai2016coupled}. 
We verified our algorithm against the analytical result of a tilted grain growing in a temperature gradient, i.e. specifically reproducing the results of Fig.~2 in Ref.~\cite{gandin19973d}.

The growth velocity of the square vertices, $V$, was estimated using 

\begin{align}
\label{eq:kgt}
V_{\rm KGT} = 6.45\times10^{-8} \times \Delta T ^ {3.83} + 5.71\times10^{-6} \times \Delta T^ {1.98},
\end{align}
where the undercooling $\Delta T=T_L-T$ is measured at the center of the cell.
The coefficients and exponents of Eq.~\eqref{eq:kgt} were adjusted to approximate the results of the classical Kurz-Giovanola-Trivedi model \cite{kurz1986theory} at $V<0.1~$m/s using parameters of the Ni-5wt\%Nb considered in the PF simulations.

For simplicity, all CA simulations consider a frame fixed to the material and an evolving temperature field following either Eq.~\eqref{eq:Tlong} with $x_0(t)=x_0(0)+V_{\rm L}\times t$ or a shrinking melt pool following Eq.~\eqref{eq:Tcross}.
All other alloy properties, processing parameters, and boundary conditions are similar to that used in the PF simulations.

The selected CA algorithm has two major computational advantages over the PF method: (1) the CA model relies exclusively on the temperature field and does not require tracking of the solute concentration field, and (2) the individual cell size $h_{\rm CA}$ used in the CA can be chosen to be much coarser than that required for accurate PF simulations (typically an order of magnitude smaller than the dendrite tip radius \cite{echebarria2004quantitative, shibuta2015solidification}). 
However, as mentioned in Section~\ref{sec:intro}, strategies for the appropriate choice of numerical parameter $h_{\rm CA}$ remain unclear.
Hence, here we aim at addressing two key questions:
\begin{itemize}
\item To what extent can a CA reproduce PF predictions of grain growth competition in a polycrystalline unsteady 2D temperature field?
\item What is an appropriate element size for the CA grid, or otherwise stated, above which value of $h_{\rm CA}$ do CA predictions start deviating significantly from reference results at low $h_{\rm CA}$?
\end{itemize}
Therefore, for each configuration (one longitudinal, fifteen cross-sections), we perform CA simulations using different $h_{\rm CA}$ and quantitatively compare the resulting grain maps with one another and with PF results (see Section~\ref{sec:method:postproc}).
We explore values of $h_{\rm CA}$ with a ratio $h_{\rm CA}/h_{\rm PF}$ between~6 and~1152 for the longitudinal section ($h_{\rm PF}=5$~nm) and between~3 and~144 for the cross-section ($h_{\rm PF}=10$~nm) simulations.

\subsection{Post-processing}
\label{sec:method:postproc}

We seek a quantitative comparison of CA and PF results on the basis of resulting grain maps and overall grain textures. 
We also want to explore the potential transition of favored grain texture in the melt pool as a function of the melt pool shape. 

In order to compare PF and CA grain maps, since we use a cell size $h_{\rm CA}$ that is systematically a multiple of the PF grid element size $h_{\rm PF}$ we can artificially refine the CA results to a grid similar to that of the PF.
To do so, we attribute the grain orientation of a given cell to all points of the refined grid included within each CA cell.
We can then compare the two grids of similar size by counting the number of points within the melt pool that have a similar or different grain orientation.

We define a region of measurement that includes only the melted region, in order to discard the unmelted Voronoi grain distribution at the bottom or sides of the domain. 
In the longitudinal simulation, we also exclude the upper region of the domain, which may be affected by the boundary conditions, namely excluding the upper $10\,\mu$m at the top of the simulation from the measurement region. 
Within the resulting region, denoted $\Gamma$, we count the number of grid point $(i,j)\in\Gamma$ in which the grain orientation differs between CA and PF, which we then normalize by the total number of grid points in $\Gamma$.
This way, we obtain a difference metric (in area fraction) that is equal to~0 if the grain maps are exactly the same and~1 if all grid points differ in the considered region $\Gamma$.

We also extract overall grain orientation distribution within each melt pool, within the same melted region~$\Gamma$.
We only perform this analysis on the cross-section simulations, which present a higher degree of statistical relevance with five different simulations performed for each configuration. 
The grain distributions are simply extracted as the number of grid points of a given orientation, combined over the five runs for each configuration, and then binned into orientation ranges of width $9\degree$ and centered on $0\degree$, $9\degree$, $18\degree$\dots $90\degree$ (the last bin centered on $90\degree$ being the same as that centered on $0\degree$).

In order to correlate the grid element sizes to physical length scales of the microstructure, we also extract values of primary and secondary arm spacings within the PF-simulated microstructures.
To do so, we perform line scans on the solute composition field (similar to those presented in Fig.~9 of Ref.~\cite{elahi2022multiscale}, where scans A, B, C, and E correspond to primary spacings and scan D corresponds to secondary spacings) in large grains presenting the clearest primary or secondary spacings.
From there, the estimation of the spacing is straightforward from locating the first and last concentration peaks (corresponding to interdendritic regions) and counting the number of spacings in between.
The resulting average spacings correspond to a total of 63 primary and 159 secondary measured spacings in the longitudinal simulation, and a total of 196 primary and 388 secondary measured spacings in the cross-section simulations.
We did not estimate secondary spacings in the shallow melt pools that were composed principally of primary dendrites, and we averaged the primary ($\lambda_1$) and secondary ($\lambda_2$) spacings over the different melt pools sizes, as we did not see any significant differences between melt pool shapes.

%=================================================================

\section{Results and discussions}

\subsection{Computational performance}
\label{sec:perform}

Phase-field simulations were parallelized on multiple graphics processing units (GPUs)~\cite{elahi2022multiscale}.
The resulting longitudinal section simulation had over one billion grid points ($50\,000\times20\,000$) iterated over 6.6 million time steps, and was performed in less than 10 days using eight Nvidia RTX 2080Ti GPUs. 
The cross-section simulations considered different melt pool sizes (see Section~\ref{sec:method:configs}) and hence different grid sizes ($12\,900\times19\,300$, $25\,700\times9\,700$, and $51\,300\times4\,900$).
Each of them has approximately 250 million grid points iterated over 858 thousand time steps, corresponding to a physical simulated time of 2.6\,ms.
The simulations were performed on various Nvidia GPU models, most of them using two RTX3090 GPUs.
All simulations performed with two RTX3090 GPUs were completed in less than 44h (wall time).
The observed strong scaling with the number of GPUs is nearly linear ($\pm10\%$).

As expected, the coarser CA grid relative to PF simulations leads to a substantial computational advantage.
Instead of using multiple parallel GPUs as in PF simulations, each CA run was performed on a single Intel Xeon 6130 CPU core (2.10\,GHz).
For the longitudinal section simulation, compared to the 10 days using eight GPUs (PF), the CA simulations were completed in 30 seconds or less (for $h_{\rm CA}/h_{\rm PF}\geq192$) and up to 10.5 days (for $h_{\rm CA}/h_{\rm PF}=6$).
Typical simulations with an intermediate grid, e.g. with $h_{\rm CA}/h_{\rm PF}=24$, last about 4 hours.
As for the cross-section simulation, compared to the 44h using two GPUs (PF), the CA simulations with the reference melt pool size were completed in between a few seconds ($h_{\rm CA}/h_{\rm PF}=144$) and 4 days ($h_{\rm CA}/h_{\rm PF}=3$).
Simulations with an intermediate grid, with $h_{\rm CA}/h_{\rm PF}=24$, last about 11 minutes.
Computational times for all configurations are listed in the joint Supplementary Material (Section~C).

\subsection{Longitudinal section}
\label{sec:longitud}

\begin{figure}[b!]
    \centering
    \includegraphics[width=3in]{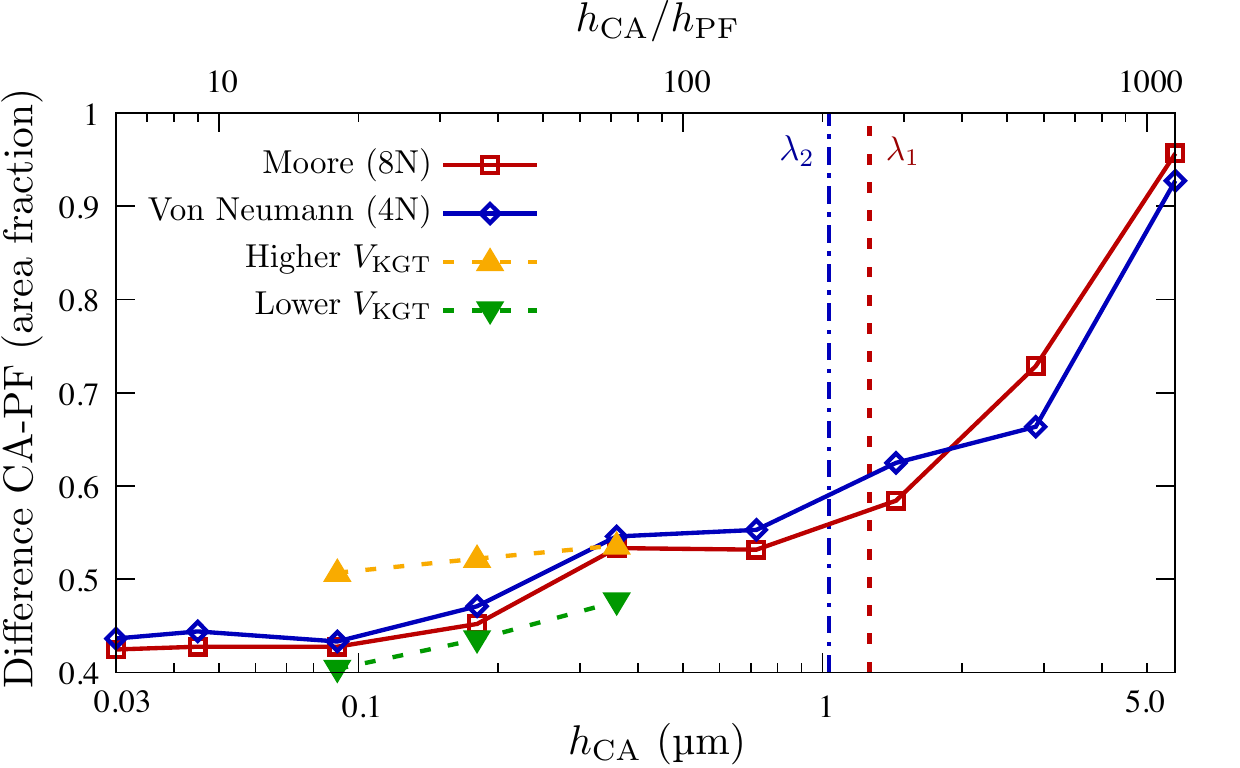}
    \caption{
    Difference between orientation maps in CA and PF results, as a fraction of the measured area $\Gamma$, considering CA results using either Moore (eight-neighbor) or von Neumann (four-neighbor) neighborhood, as well as artificially higher ($V=10\times V_{\rm KGT}$) or lower ($V=V_{\rm KGT}/10$) growth velocities, i.e. respectively multiplying or dividing the right-hand side of Eq.~\eqref{eq:kgt} by a factor of 10. Vertical lines show the measured values of primary ($\lambda_1$) and secondary ($\lambda_2$) dendrite arm spacings.
    }
    \label{fig3}
\end{figure}

Figure~\ref{fig3} shows the difference, in area fraction, between CA and PF grain maps, in the measured area $\Gamma$, for different CA grid sizes.
The measured primary and secondary arms spacings, respectively $\lambda_1=1.26~\mu$m and $\lambda_2=1.03~\mu$m, are also reported as vertical lines.

%%%%%%%%%%%%%%%%
%%%%%%%%%%%%%%%%

Several important observations emerge from Fig.~\ref{fig3}.
First, the difference between CA and PF is essentially increasing as the CA grid gets coarser, and the CA results seem to converge to a plateau as $h_{\rm CA}$ is reduced. 
However, even the best results for a well-converged CA, reached for approximately $h_{\rm CA}\leq0.2~\mu$m (i.e. $h_{\rm CA}/h_{\rm PF}\leq40$), still show over 40\% of difference with PF.
This $h_{\rm CA}$ is significantly lower than the measured dendrite arm spacings $\lambda_1=1.26~\mu$m and $\lambda_2=1.03~\mu$m.
This monotonic behavior contrasts with previous results in bi-crystal configurations \cite{pineau2018growth,dorari2022growth}, which have shown that a minimum can be found somewhere between the so-called \emph{geometrical limit} (at low $h_{\rm CA}$) and the \emph{favorably-oriented grain} (at high $h_{\rm CA}$) asymptotic behaviors.
At the highest $h_{\rm CA}$, the CA prediction fails over more than 90\% of the solidified area, as the cell size becomes comparable to the grain size. 
From these results, while the most appropriate choice for the CA grid size seems to be toward low  $h_{\rm CA}$, which leads lower difference with PF, the correlation with microstructural length scales remains unclear and requests further investigation.

We also explored the effect of some key features of the CA model on these results, such as the neighborhood considered for cell captures, as well as the growth law considered for the dendrite tips (i.e. square vertices).
Changing the capture neighboring algorithm from Moore (eight neighbors, used as default here) to von Neumann (four neighbors) definition does not affect the results substantially. 

Regarding the growth velocity, we find that increasing the growth velocity (by multiplying $V$ from Eq.~\eqref{eq:kgt} by a factor 10) or decreasing it (dividing by a factor 10) has a relatively limited effect, at least in this specific example.
However, such conclusion needs to be taken carefully, keeping in mind that we are here in presence of a relatively high temperature gradient. 
One main effect of a higher $V(\Delta T)$ is that the solidification front (dendrite tips) will be located closer to the liquidus temperature, stabilizing at a relatively lower $\Delta T$ for the same $V$, while a lower  $V(\Delta T)$ will promote a stabilization of the solidification front deeper in the mushy zone (closer to the solidus temperature).
Due to the high temperature gradient considered here, between $10^{6}$ at the bottom and $10^{7}$\,K/m at the tail of the melt pool (See Ref.~\cite{elahi2022multiscale}, Fig.~5 therein), the solidus and liquidus isotherms are relatively close to each other, which minimizes the effect of $V(\Delta T)$.

\begin{figure}[b!]
    \centering
    \includegraphics[width=3in]{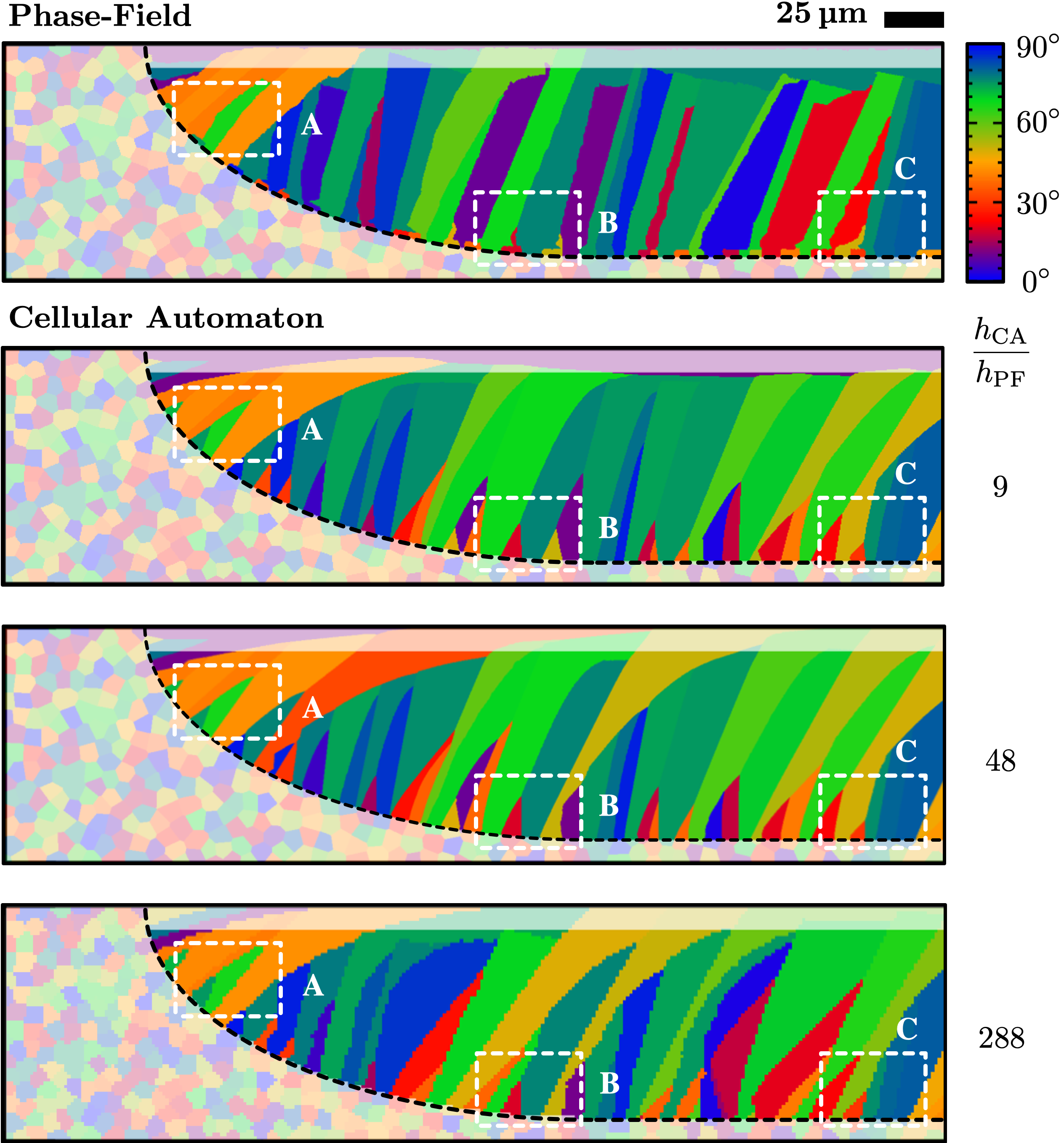}
    \caption{Selected grain orientations (color map) predicted by phase-field (top row) and cellular automaton (bottom rows) with different grid coarsening levels $h_{\rm CA}/h_{\rm PF}=9$, 48, and 288. 
The brighter (washed out) area is outside of the region $\Gamma$ considered for the quantitative measurements and comparisons of grain distributions.
The dashed black line represents the limit of the melted region.
    }
    \label{fig1}
\end{figure}

Figure~\ref{fig1} shows the resulting grain maps in the longitudinal section for the well-converged ($h_{\rm PF}=5\,$nm) PF simulation, as well as for three different levels of CA grid size, namely $h_{\rm CA}/h_{\rm PF}=9$, 48, and 288.
We did not analyze quantitatively the average grain distributions in the melt pool, because the number of grains in this single simulation does not provide a sufficient sample size for statistically meaningful results.
Still, the overall distribution of grains appears close in most simulations, with a grain orientation, i.e. a preferred growth direction $\alpha$ measured counterclockwise from the horizontal $x+$ direction, close to $60\degree$ (predominantly shades of green) in most of the melt pool, and closer to $45\degree$ (orange) around the tail of the melt pool. 
However, the PF simulation shows a broader distribution of orientations with some large grains at $\alpha\approx15\degree$ (purple), $30\degree$ (red), and $0\degree$ (blue), which are much less present in CA-predicted grain maps.

%%%%%%%%%%%%%%%%
%%%%%%%%%%%%%%%%

\begin{figure}[b!]
    \centering
    \includegraphics[width=3in]{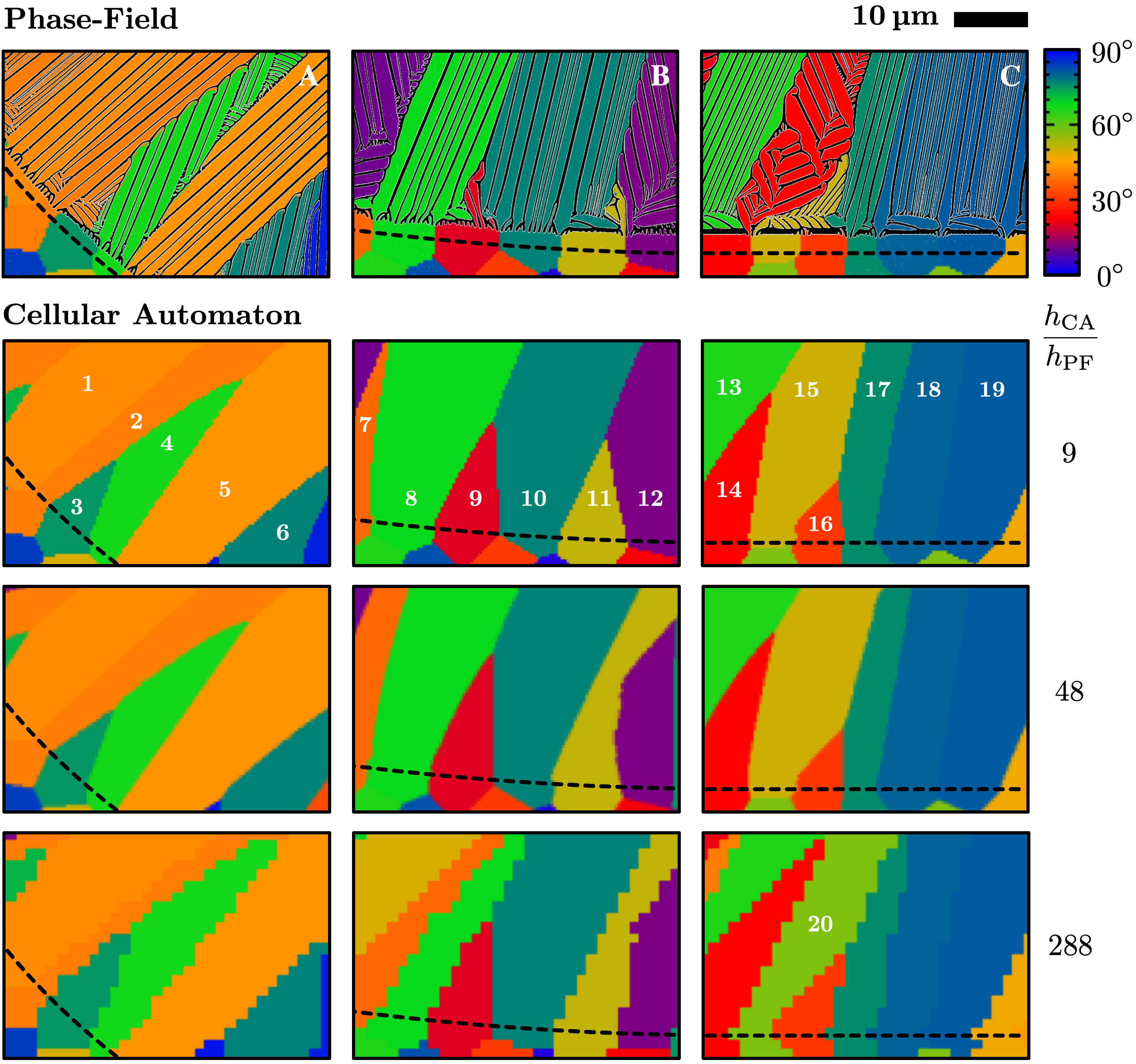}
    \caption{Comparison of selected grain orientations (color map) predicted by phase-field (top row) and cellular automaton (bottom rows) with different grid coarsening levels $h_{\rm CA}/h_{\rm PF}=9$, 48, and 288, within the regions A, B, and C highlighted in Fig.~\ref{fig1}.
    The PF grain map is overlaid with the inner grain dendritic structure.
    The dashed black lines show the limit of the melted region.
    }
    \label{fig2}
\end{figure}

Figure~\ref{fig2} provides a more detailed view of three regions labelled A, B, and C in Fig.~\ref{fig1}.
The underlying dendritic structures overlaid on the PF grain map allows to identify the mechanism of grain growth competition at play at each GB (namely impingement at converging GBs and sidebranching at diverging GBs \cite{tourret2015growth}) and to better understand the agreements and differences between PF and CA predictions.

Region A, located close to the tail of the melt pool, illustrates a region of good agreement between PF and CA, nearly regardless of the CA grid. 
Most grain boundaries are of converging type, except for diverging GB $4-5$.
Considering that the main direction of the temperature gradient is normal to the fusion line (marked in dashed black line), the preferred growth orientation is close to $\alpha\approx45\degree$.
As a result, all converging GBs in this region follows the favorably-oriented grains.
The orientation of the diverging GB $4-5$ is between that of grains 4 and 5, with sidebranches emerging from both grains, i.e. close to the geometrical limit.
The only notable deviation between PF and CA results is the early disappearance of grain 3 in PF results, due to early sidebranching of favorably-oriented grain 2.

Region B, at the bottom of the melt pool, contains converging ($7-8$, $8-9$, $8-10$, $10-11$, $10-12$) and diverging ($9-10$, $11-12$) GBs.
In PF simulations, sidebranching at diverging GBs and at converging GBs $8-9$ and $10-11$ results in early elimination of the least-favorably oriented grains 9 and 11 (considering a temperature gradient essentially vertical with a slight tilt toward the right).
These grains survive longer in CA simulations.
For coarse grids (bottom rows), grain 11 even survives to become one of the major grains in the final microstructure (see Fig.~\ref{fig1}) as the temperature gradient direction tilts closer to its preferred growth direction ($\alpha\approx45\degree$) when the melt pool advances.
Another notable difference between PF and CA is the absence in the PF map of grain $7$, also victim of early sidebranching from neighboring grains.

%%%%%%%%%%%%%%%%
%%%%%%%%%%%%%%%%

Region C, also at the bottom of the melt pool, presents important discrepancies between the predicted maps.
There, converging GBs are $15-16$ and $16-17$, and diverging GBs are $13-15$ and $14-15$.
Region C also contains the low-angle GBs $17-18$ and $18-19$.
Other grain boundaries involving grain 14 (i.e. $13-14$ and $14-17$, the latter existing only in the PF simulation) are not clearly defined, as the misorientation with its neighbors is close to $45\degree$.
In a static temperature gradient, a misorientation of $45\degree$ with the temperature gradient would have lead to the growth of a degenerate (seaweed) microstructure growing at a much higher undercooling and strongly penalized in the grain growth competition \cite{tourret2015growth}.
However, here, grain 14 survives for a long time in PF simulations (see Fig.~\ref{fig1}) in spite of the strong early competition with neighbor grain 13.
Here too, a major discrepancy between PF and CA stems form the early elimination of grain 15 in PF, which does not occur in CA simulation at low $h_{\rm CA}$, even leading to the grain becoming one of the major grains in the subsequent grain growth competition (see Fig.~\ref{fig1}).
The coarsest CA simulation (bottom row) even leads to a completely different outcome for grain 15. 
The initial GB between grains 15 and 20 being close to the fusion line (black dashed line), a coarse discretization with $h_{\rm CA}/h_{\rm PF}\geq288$ leads to the complete melting of grain 15, hence suppressed and replaced by lower grain 20. 
It follows that, as the spatial discretization gets coarser, the identity and orientation of initially competing grains may be completely changed.

A notable difference between PF and CA simulations is the presence of a region just above the fusion line that, in PF simulations, initially grows with a planar solid-liquid interface before the morphological destabilization into dendrites, after which the actual growth competition among dendrites starts.
Meanwhile, in CA simulations the grains start competing from the very onset of the simulation, and the main orientation of GBs (mostly linear at this scale) is initiated directly from the fusion line.
These differences between CA and PF, also observed in all cross-section simulations presented later, highlight the fact that morphological transitions (e.g. planar to dendrite) and the prediction of non-planar (rough) GBs requires the level of accuracy provided by PF.

\subsection{Cross-sections}
\label{sec:cross}

\subsubsection{Effect of CA grid}
\label{sec:cross:grid}

%%%%%%%%%%%%%%%%
\begin{figure}[b!]
    \centering
    \includegraphics[width=3in]{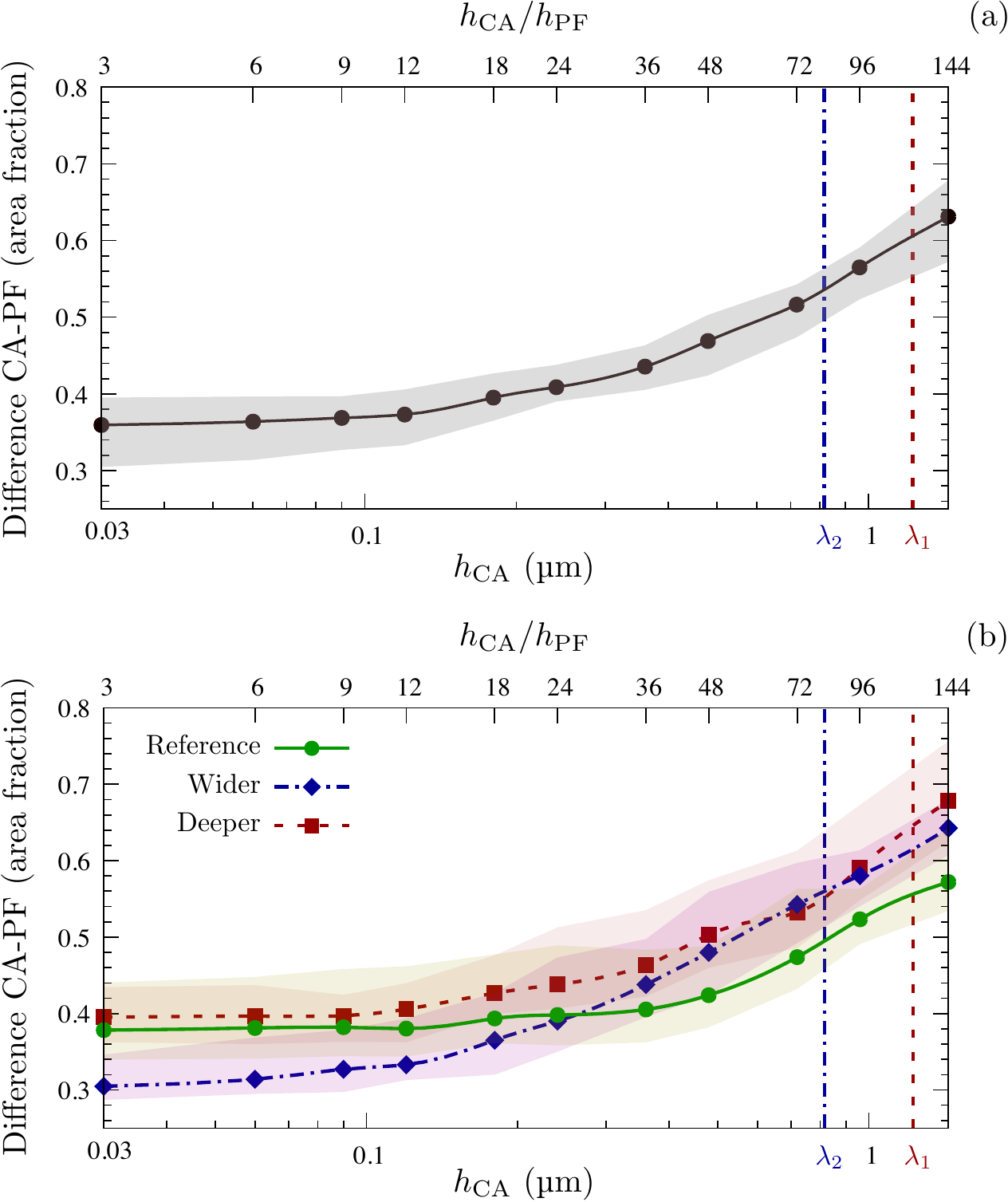}
    \caption{
    Difference between orientation maps in CA and PF results, as a fraction area of the measured area $\Gamma$, considering CA results in the cross-section simulations: (a) averaged over all 15 simulations, (b) averaged over five simulations for each melt pool shape (reference, wider/shallower, and deeper/narrower).
    The shaded area associated with each curve represents the distribution between the five different runs (minimum and maximum of CA-PF difference).
    }
    \label{fig4}
\end{figure}
%%%%%%%%%%%%%%%%

Figure~\ref{fig4} shows the difference between CA and PF grain maps in the measured areas of the cross-section simulations for different CA grid sizes: (a) averaged over all 15 simulations or (b) averaged over the five simulations for each melt pool shape.
In these cases, the primary and secondary dendrite arms spacings did not show any significant dependence on the melt pool shape, except for the wide and shallow melt pools that essentially exhibited only primary dendrites with very few secondary branches.
Hence, we averaged the spacings over all measurable simulations, yielding $\lambda_1=1.22~\mu$m and $\lambda_2=0.82~\mu$m 

While the errors at low $h_{\rm CA}$, between 30\% and 40\%, appear slightly lower than in the longitudinal simulation, the observed trends are essentially similar. 
Namely, the difference between PF and CA maps increases when  $h_{\rm CA}$ increases, and the low-$h_{\rm CA}$ plateau is reached at a grid size much smaller than the characteristic microstructural length scales $\lambda_1$ and $\lambda_2$.
In spite this apparent poor PF-CA match, we show in the following subsection that, even though the discrepancy remains high when integrating the difference point-by-point, from a statistical point of view, disregarding the exact location or shape of grains, the finer CA simulations provide a good representation of the average statistical distribution of grain orientations.

\subsubsection{Effect of melt pool shape}
\label{sec:cross:shape}

\begin{figure}[b!]
    \centering
    \includegraphics[width=3in]{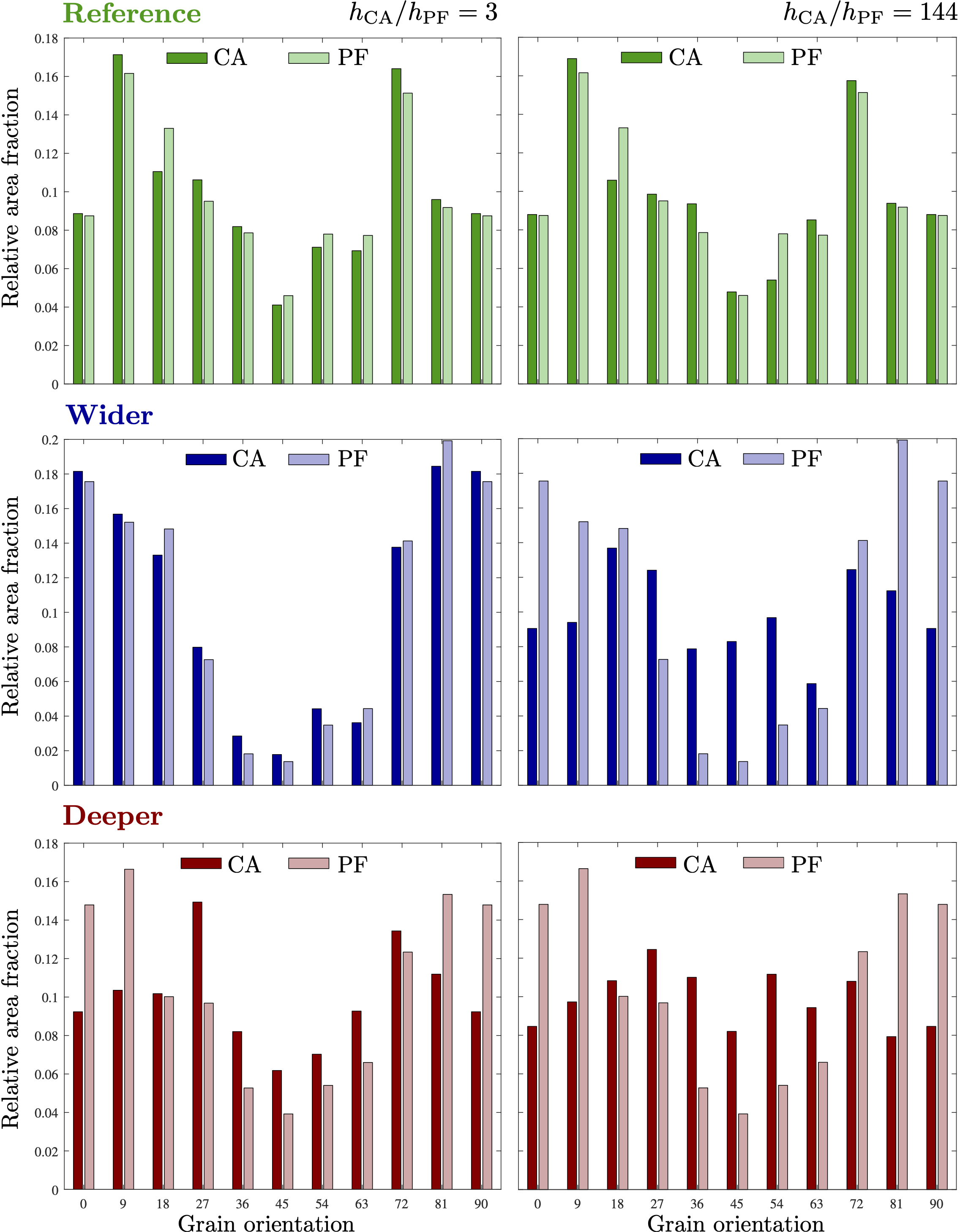}
    \caption{
    Distribution of grain orientations, averaged over five simulations for three different melt pool shapes, namely the reference case fitted to FE results (top row), a wider and shallower melt pool (central row), and a deeper and narrower melt pool (bottom row), comparing PF and CA results for a fine ($h_{\rm CA}/h_{\rm PF}=3$, left column) and a coarse ($h_{\rm CA}/h_{\rm PF}=144$, right column) CA grid. (PF results are the same in left and right columns, only duplicated for readability.)
    }
    \label{fig5}
\end{figure}

Figure~\ref{fig5} shows the histograms of grain distributions averaged over five simulations for the three different melt pool shapes (rows), comparing PF simulation results (lighter color) with CA predictions (darker color) with fine ($h_{\rm CA}/h_{\rm PF}=3$, left) and coarse ($h_{\rm CA}/h_{\rm PF}=144$, right) grids.

%%%%%%%%%%%%%%%%
%%%%%%%%%%%%%%%%

Simple geometrical considerations \cite{jadhav2019influence} suggest that, due to the main temperature gradient direction, a transition occurs between a predominant $\langle100\rangle$ grain texture in the horizontal built direction (i.e., in 2D, a high density of grains with $\alpha\approx0\degree$) for a relatively wide/shallow or a deep/narrow melt pool to a predominant $\langle110\rangle$ texture (i.e. $\alpha\approx45\degree$) for melt pool cross-section aspect ratio closer to unity (i.e. close to circular in shape). 
This trend was supported by experimental observations, for instance from powder-bed laser melting of pure copper \cite{jadhav2019influence} and pure molybdenum \cite{higashi2020selective}.
While the total number of simulations, and hence grains, presented here is not statistically sufficient to capture such a transition of grain texture, Fig.~\ref{fig5} exhibits clear trends in this direction, considering PF results as the most accurate ones.
In particular, the wide and shallow melt pools (central row) markedly exhibit a low density of $\langle11\rangle$ ($\alpha\approx45\degree$) oriented grains and hence a stronger $\langle10\rangle$ ($\alpha\approx0\degree$) texture compared to reference (top row) and deeper/narrower (bottom row) melt pools.

\begin{figure}[b!]
    \centering
    \includegraphics[width=3in]{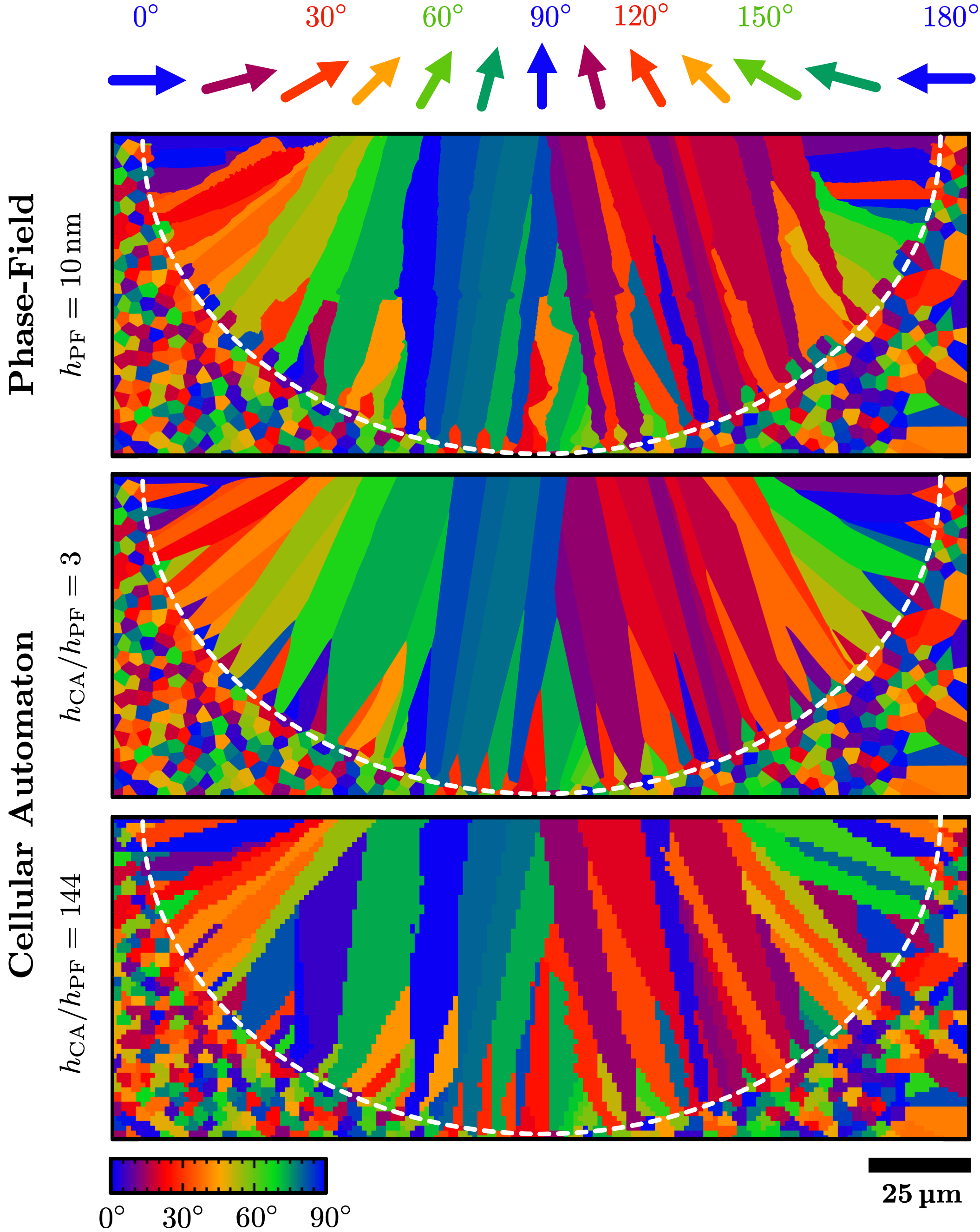}
    \caption{
        Grain orientation maps for the reference melt pool dimension given by FE thermal simulation ($w_0=128~\mu$m, $d_0=96~\mu$m), predicted by phase field (top row) compared to CA with fine (central row) and coarse (bottom row) grid.
    The white dashed lines identify the melted region.
    Arrows at the top illustrate the predominant selected grain orientations within the melt pool.
    }
    \label{fig6}
\end{figure}

For the reference melt pool shape (Fig.~\ref{fig5}, top row), which has the aspect ratio closest to unity ($w_0/d_0=1.33$), the agreement between PF and CA is overall excellent, regardless of the CA grid size.
Figure~\ref{fig6} shows the typical grain distribution in such a melt pool.
In this simulation, the selected grain orientation clearly follows the main temperature gradient direction across the melt pool.
This results in a $\langle10\rangle$ texture (expressed with respect to the build direction, here vertical) on the sides and center of the melt pool, with intermediate orientations smoothly changing along the width of the melt pool.
This trend is well captured by both CA simulations.
Overall, the key differences between PF and CA simulations are: (1) the presence of the early planar growth region just above the fusion line in PF simulation, (2) the early elimination of least favorably oriented grains in PF appearing more progressively in CA simulations (as described in Section~\ref{sec:longitud}), and (3) the markedly higher roughness of grain boundaries in the PF simulations, while all GBs from CA simulations are essentially straight lines.
The simulation illustrated in Fig.~\ref{fig6} and this characteristic distribution of grain orientations within the melt pool is representative of the trend observed in all five simulations (see all of them in Supplementary Material -- Fig.\,B.1).

%%%%%%%%%%%%%%%%
%%%%%%%%%%%%%%%%

%%%%%%%%%%%%%%%%
\begin{figure*}[t!]
    \centering
    \includegraphics[width=6in]{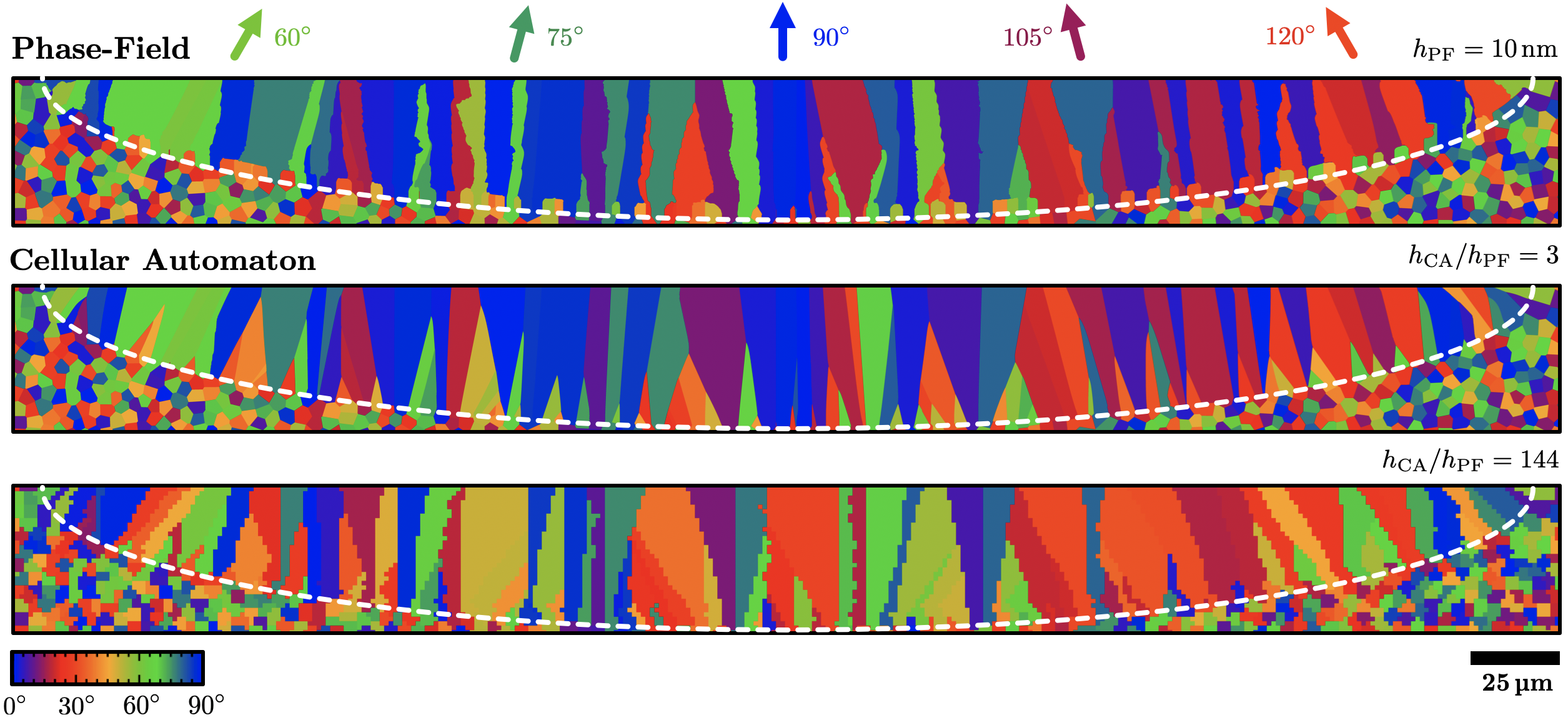}
    \caption{
        Grain orientation maps for a wide and shallow melt pool ($w_0=256~\mu$m, $d_0=48~\mu$m), predicted by phase field (top row) compared to CA with fine (central row) and coarse (bottom row) grid.
    The white dashed lines identify the melted region.
    Arrows at the top illustrate the predominant selected grain orientations within the melt pool.
    }
    \label{fig7}
\end{figure*}
%%%%%%%%%%%%%%%%

%%%%%%%%%%%%%%%%
\begin{figure*}[t!]
    \centering
    \includegraphics[width=4.5in]{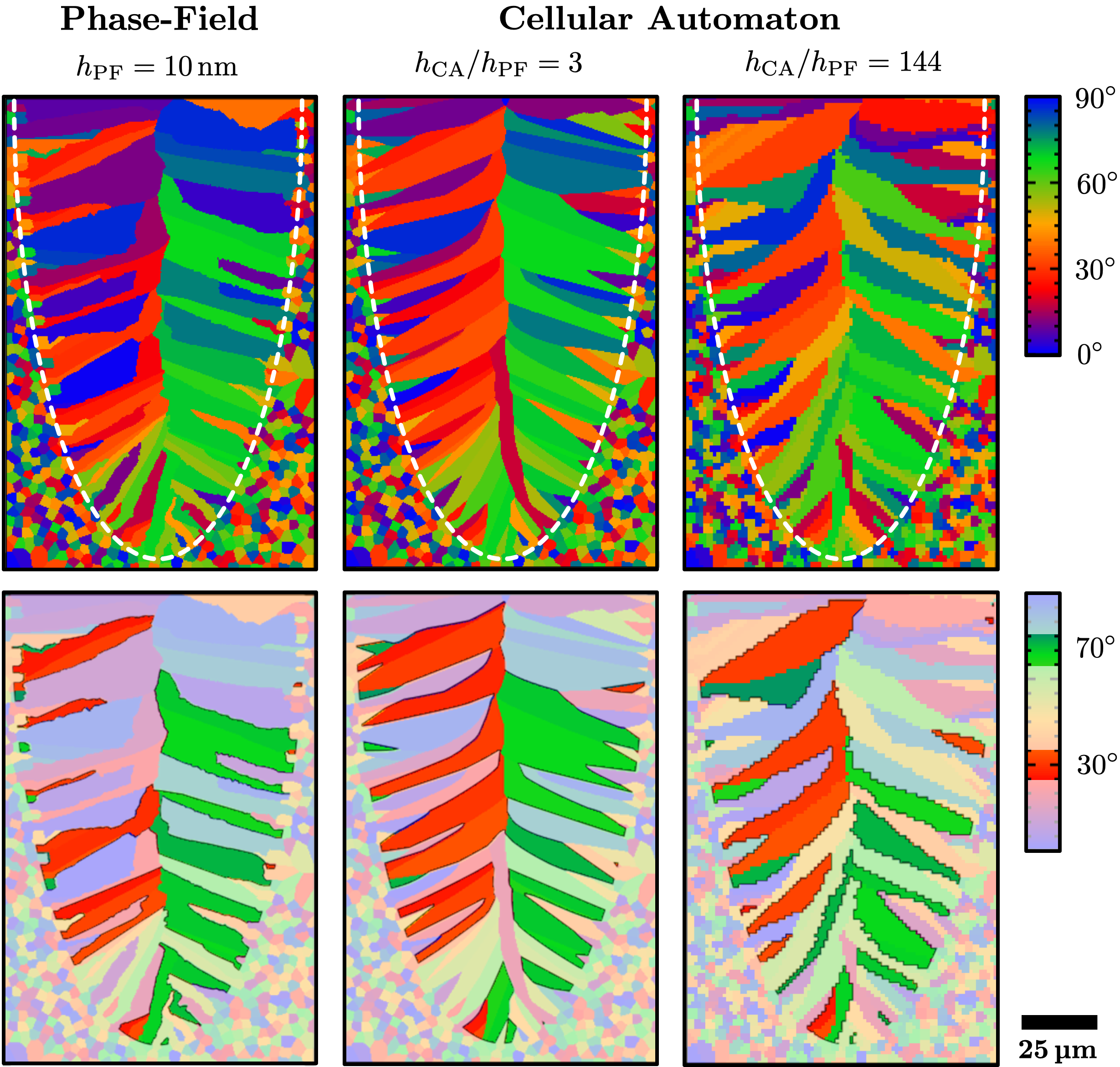}
    \caption{
    Grain orientation maps for a deep and narrow melt pool ($w_0=64~\mu$m, $d_0=192~\mu$m), predicted by phase field (left column) compared to CA with fine (central column) and coarse (right column) grid.
    The white dashed lines identify the melted region.
    The bottom row highlights grains with orientations $25\degree\leq\alpha\leq35\degree$ (red) and $65\degree\leq\alpha\leq75\degree$ (green).
    }
    \label{fig8}
\end{figure*}
%%%%%%%%%%%%%%%%

In the wider (i.e. shallower) melt pool (Fig.~\ref{fig5}, central row), with an aspect $w_0/d_0=5.33$, the agreement between PF and CA is excellent for fine CA grids (left column), but a notable discrepancy appears for coarser CA grids (right column).
Figure~\ref{fig7} illustrates the selected grain maps for a given initial grain distribution, representative of the behavior observed in all five configurations (see all of them Supplementary Material -- Fig.\,B.2)
In this case, the temperature gradient is essentially oriented upwards.
Therefore, it is not surprising to observe that in the PF simulations, the vast majority of selected grains have their orientation within $\alpha=90\degree\pm30\degree$.
For a coarser CA grid (bottom row), the distribution of grain orientations across the melt pool is more scattered, and several grain with $\alpha\approx45\degree$ even appear within the solidified melt pool.
Similarly as in previously described cases, the initial planar growth followed by early elimination of unfavorably oriented grains (resulting in little square grains just above the fusion line) is observed only in PF simulations, which also exhibits notably rougher GBs.

For the deeper (i.e. narrower) melt pools (Fig.~\ref{fig5}, bottom row), the agreement between CA and PF grain orientation densities is notably worse, and it only marginally improves with lower $h_{\rm CA}$.
However, in spite of a relatively higher discrepancy in their relative amount, Figure~\ref{fig8} shows that both PF and CA qualitatively match reasonably well in terms of texture distribution within the melt pool, commensurately with the main temperature gradient direction therein.
The left-hand side of the melt pool is mostly composed of grains with orientation $0\degree\leq\alpha\leq35\degree$, while the right-hand side exhibits a high density of grain with $55\degree\leq\alpha\leq90\degree$.
This left-right segregation of grain orientations is highlighted in the bottom part of Fig.~\ref{fig8}, where orientations are highlighted that fall between $25\degree$ and $35\degree$ (red) and between $65\degree$ and $75\degree$ (green).
Compared to the reference and wide melt pool cases, in which CA-predicted GBs are essentially straight lines, those in the deeper melt pool are more curved, as they follow the rotation of the main temperature gradient direction during the melt pool solidification.
In the PF simulation, this overall rotation of the grain from horizontal to upwards also appears, but the morphologies of GBs are, once again, rougher than in CA simulations.
This behavior is representative of observations in all five simulated deep melt pools (see Supplementary Material -- Fig.\,B.3).

%=================================================================

\section{Summary, Conclusions, and Perspectives}

In this article, we performed systematic comparisons between grain microstructures selected during the polycrystalline solidification of a melt pool predicted by phase-field and cellular automaton techniques.
We used a binary Ni-5wt\%Nb as model alloy, and a thermal field calculated by finite elements thermal simulations for an Inconel~718 alloy \cite{elahi2022multiscale}, and analyzed 2D simulations of solidification in the longitudinal direction \cite{elahi2022multiscale}, as well as different melt pool shapes in the cross-section normal to the scanning direction.

Our simulations revealed that the exact matching of grain structures, i.e. comparing detailed grain and GB morphologies, essentially improves with the refinement of the CA grid.
When comparing/integrating grain orientation maps point-by-point, the best CA simulations capture the PF grain maps over only 70\% of the melted region at best.
However, while specific details at the scale of individual grains and GBs are not accurately reproduced, the overall statistical distribution of crystal orientations in the entire melt pools, averaged over several runs, is relatively well predicted by CA simulations using finer grids, for a fraction of the PF computational cost.
This agreement seems to depend also on the melt pool shape, as we found poorer matching for the deeper melt pools -- which we attribute to the stronger change in temperature profile and gradient directions during solidification.

Looking at the detailed GB scale, the trajectories of converging and low angle GBs are most often (yet not always) well captured, as most of them tend to follow the classical favorably-oriented grain criterion.
The discrepancy is higher for high-angle diverging GBs.

Some striking differences appear between PF and CA related to transient growth regime and solid-liquid interface morphology.
Indeed, PF simulations systematically lead to an initial transient solidification period with a nearly planar solid-liquid interface, until its velocity increases enough to lead to a morphological destabilization into cells or dendrites.
This is not captured by CA simulations, in which the grains start competing directly from the onset of the simulation (i.e. from the fusion line).
In PF simulations, the transient destabilization of the planar interface into dendrites also gives rise to intense sidebranching activity, from which the least favorably oriented grains (i.e. those with higher misorientation of their preferred growth direction with the temperature gradient) are often rapidly eliminated by favorably oriented neighbors.
This dynamics of sidebranching following the morphological destabilization of the interface is not captured by CA simulations.
Moreover, the CA-predicted GBs are much smoother than the rough ones emerging from PF simulations.

Regarding the effect of the melt pool shape, melt pools with an aspect ratio close to unity (i.e. near circular) exhibit a good agreement between CA and PF statistical distribution of grain orientations, as well as their distribution within the melt pool.
The CA-PF agreement becomes lower, and more grid-dependent, for higher aspect ratio of the melt pool, while the overall distribution of grain orientations across the melt pool remains reasonable.
Our 15 configurations did not provide enough statistics to unambiguously capture the transition from $\langle100\rangle$ to $\langle110\rangle$ as a function of melt pool aspect ratio \cite{jadhav2019influence}.
However, PF, and to some extent fine-grid CA simulations, show clear indications that such experimentally-observed transition \cite{jadhav2019influence,higashi2020selective} can be predicted, should the number of simulations (and hence grains) be increased to a more statistically significant amount.

Furthermore, our simulations did not exhibit a strong dependence upon the considered CA neighborhood algorithm (Moore or von Neumann) nor upon the dendrite growth kinetics law $V(\Delta T)$ applied at the grain vertices.
We attribute the latter to the high temperature gradient, reducing the distance between solidus and liquidus isotherms, and hence reducing the effect of the growth law upon the location of the solidification front between $T_L$ and $T_S$.

Some limitations of the current study are worth highlighting, several of which are actually the focus of ongoing work building up from this study.
First, solidification simulations presented here, while based on 3D calculation of the temperature field, are two-dimensional, thus not capturing the full complexity of grain growth competition in an actual melt pool. 
However, the present study would, at present, not be possible in 3D, as the grid size required by well-converged PF simulations already leads to sizable 2D simulations, handled here with an advanced multi-GPU parallelized implementation.
Other important phenomena present during melt pool solidification were ignored, such as fluid flow and nucleation. 
The latter would specifically deserve a closer look, as nucleation parameters (e.g. undercooling and site density) are almost systematically taken as calibration parameters in both PF and CA simulations.
Finally, more advanced CA formulations, e.g. using irregular quadrangles instead of square building blocks \cite{carozzani2012developpement, pineau2018growth, fleurisson2022hybrid}, could also have a key influence, considering the high temperature gradients encountered in AM-relevant conditions.

In spite of these limitations, we trust that the presented results bring some useful insight into the applicability and limitations of both modeling methods.
They clearly show that, when considering microstructural details at the scale of individual grains or GBs, the accuracy of PF remains required to capture transient growth kinetics and solid-liquid interface morphological evolution.
The prediction of transient growth regimes in CA simulations might be improved using more advanced growth kinetics \cite{fleurisson2022hybrid} instead of the classical Ivantsov- or KGT-based ones.
However, the modeling of the morphological stability of a solid-liquid interface unambiguously requires the level of details provided only by PF.
This is even more important in AM-relevant conditions, that may lead to the restabilization of a planar interface at high growth velocity, called {\it absolute stability} \cite{mullins1964stability, kurz1994rapid}, and evidenced in several rapidly solidified alloys in the form of banded microstructures \cite{carrard1992banded, kurz1996banded, mckeown2016time}.
Still, when the most important length scale is that of the average grain microstructure, e.g. to predict averaged anisotropic mechanical properties over representative volume elements \cite{pilgar2022effect}, our results show that CA-based methods are capable of statistically reasonable grain texture predictions.
As similar quantitative PF simulations in 3D are currently computationally out of reach, the substantial speed-up and upscaling afforded by CA modeling provides a most promising route for the prediction of actual three-dimensional AM microstructures.

%=================================================================

\section*{Acknowledgements}

This work was supported by the Spanish Ministry of Science and the European Union NextGenerationEU (PRTR) through the MiMMoSA project [PCI2021-122023-2B] and a Ram\'on y Cajal fellowship [RYC2019-028233-I].

%=================================================================

 \bibliography{References}

\begin{thebibliography}{10}
\expandafter\ifx\csname url\endcsname\relax
  \def\url#1{\texttt{#1}}\fi
\expandafter\ifx\csname urlprefix\endcsname\relax\def\urlprefix{URL }\fi
\expandafter\ifx\csname href\endcsname\relax
  \def\href#1#2{#2} \def\path#1{#1}\fi

\bibitem{carroll2015anisotropic}
B.~E. Carroll, T.~A. Palmer, A.~M. Beese, Anisotropic tensile behavior of
  ti--6al--4v components fabricated with directed energy deposition additive
  manufacturing, Acta Materialia 87 (2015) 309--320.

\bibitem{kirka2017mechanical}
M.~M. Kirka, F.~Medina, R.~Dehoff, A.~Okello, Mechanical behavior of
  post-processed inconel 718 manufactured through the electron beam melting
  process, Materials Science and Engineering: A 680 (2017) 338--346.

\bibitem{gordon2018fatigue}
J.~Gordon, C.~Haden, H.~Nied, R.~Vinci, D.~Harlow, Fatigue crack growth
  anisotropy, texture and residual stress in austenitic steel made by wire and
  arc additive manufacturing, Materials Science and Engineering: A 724 (2018)
  431--438.

\bibitem{mooney2019process}
B.~Mooney, K.~I. Kourousis, R.~Raghavendra, D.~Agius, Process phenomena
  influencing the tensile and anisotropic characteristics of additively
  manufactured maraging steel, Materials Science and Engineering: A 745 (2019)
  115--125.

\bibitem{hosseini2019review}
E.~Hosseini, V.~Popovich, A review of mechanical properties of additively
  manufactured inconel 718, Additive Manufacturing 30 (2019) 100877.

\bibitem{walton1959}
D.~Walton, B.~Chalmers, The origin of the preferred orientation in the columnar
  zone of ingots, Trans, AIME 215 (1959) 447--456.

\bibitem{dsouza2002morphological}
N.~D'souza, M.~Ardakani, A.~Wagner, B.~Shollock, M.~McLean, Morphological
  aspects of competitive grain growth during directional solidification of a
  nickel-base superalloy, cmsx4, Journal of materials science 37~(3) (2002)
  481--487.

\bibitem{wagner2004grain}
A.~Wagner, B.~Shollock, M.~McLean, Grain structure development in directional
  solidification of nickel-base superalloys, Materials Science and Engineering:
  A 374~(1-2) (2004) 270--279.

\bibitem{zhou2008mechanism}
Y.~Zhou, A.~Volek, N.~Green, Mechanism of competitive grain growth in
  directional solidification of a nickel-base superalloy, Acta Materialia
  56~(11) (2008) 2631--2637.

\bibitem{li2012phase}
J.~Li, Z.~Wang, Y.~Wang, J.~Wang, Phase-field study of competitive dendritic
  growth of converging grains during directional solidification, Acta
  Materialia 60~(4) (2012) 1478--1493.

\bibitem{tourret2015growth}
D.~Tourret, A.~Karma, Growth competition of columnar dendritic grains: A
  phase-field study, Acta Materialia 82 (2015) 64--83.

\bibitem{takaki2016two}
T.~Takaki, M.~Ohno, Y.~Shibuta, S.~Sakane, T.~Shimokawabe, T.~Aoki,
  Two-dimensional phase-field study of competitive grain growth during
  directional solidification of polycrystalline binary alloy, Journal of
  Crystal Growth 442 (2016) 14--24.

\bibitem{tourret2017grain}
D.~Tourret, Y.~Song, A.~J. Clarke, A.~Karma, Grain growth competition during
  thin-sample directional solidification of dendritic microstructures: A
  phase-field study, Acta Materialia 122 (2017) 220--235.

\bibitem{mota2015initial}
F.~Mota, N.~Bergeon, D.~Tourret, A.~Karma, R.~Trivedi, B.~Billia, Initial
  transient behavior in directional solidification of a bulk transparent model
  alloy in a cylinder, Acta Materialia 85 (2015) 362--377.

\bibitem{song2018thermal}
Y.~Song, D.~Tourret, F.~Mota, J.~Pereda, B.~Billia, N.~Bergeon, R.~Trivedi,
  A.~Karma, Thermal-field effects on interface dynamics and microstructure
  selection during alloy directional solidification, Acta Materialia 150 (2018)
  139--152.

\bibitem{mota2021effect}
F.~Mota, J.~Pereda, K.~Ji, Y.~Song, R.~Trivedi, A.~Karma, N.~Bergeon, Effect of
  sub-boundaries on primary spacing dynamics during 3d directional
  solidification conducted on declic-dsi, Acta Materialia 204 (2021) 116500.

\bibitem{liu2013dependence}
Z.~Liu, M.~Lin, D.~Yu, X.~Zhou, Y.~Gu, H.~Fu, Dependence of competitive grain
  growth on secondary dendrite orientation during directional solidification of
  a ni-based superalloy, Metallurgical and Materials Transactions A 44~(11)
  (2013) 5113--5121.

\bibitem{yu2014anomalous}
H.~Yu, J.~Li, X.~Lin, L.~Wang, W.~Huang, Anomalous overgrowth of converging
  dendrites during directional solidification, Journal of crystal growth 402
  (2014) 210--214.

\bibitem{hu2018competitive}
S.~Hu, L.~Liu, W.~Yang, J.~Zhang, T.~Huang, Y.~Wang, X.~Zhou, Competitive
  converging dendrites growth depended on dendrite spacing distribution of
  ni-based bi-crystal superalloys, Journal of Alloys and Compounds 735 (2018)
  1878--1884.

\bibitem{wang2019competitive}
Y.~Wang, S.~Li, Z.~Liu, B.~Yang, H.~Zhong, H.~Xing, Competitive growth of
  degenerate pattern and dendrites during directional solidification of a
  bicrystal metallic alloy, Metallurgical and Materials Transactions A 50~(10)
  (2019) 4677--4685.

\bibitem{zhao2001front}
P.~Zhao, J.~Heinrich, Front-tracking finite element method for dendritic
  solidification, Journal of Computational Physics 173~(2) (2001) 765--796.

\bibitem{li2003fixed}
C.-Y. Li, S.~V. Garimella, J.~E. Simpson, Fixed-grid front-tracking algorithm
  for solidification problems, part i: Method and validation, Numerical Heat
  Transfer, Part B: Fundamentals 43~(2) (2003) 117--141.

\bibitem{rodgers2017simulation}
T.~M. Rodgers, J.~D. Madison, V.~Tikare, Simulation of metal additive
  manufacturing microstructures using kinetic monte carlo, Computational
  Materials Science 135 (2017) 78--89.

\bibitem{rodgers2021simulation}
T.~M. Rodgers, D.~Moser, F.~Abdeljawad, O.~D.~U. Jackson, J.~D. Carroll, B.~H.
  Jared, D.~S. Bolintineanu, J.~A. Mitchell, J.~D. Madison, Simulation of
  powder bed metal additive manufacturing microstructures with coupled finite
  difference-monte carlo method, Additive Manufacturing 41 (2021) 101953.

\bibitem{jreidini2021orientation}
P.~Jreidini, T.~Pinomaa, J.~M. Wiezorek, J.~T. McKeown, A.~Laukkanen,
  N.~Provatas, Orientation gradients in rapidly solidified pure aluminum thin
  films: comparison of experiments and phase-field crystal simulations,
  Physical review letters 127~(20) (2021) 205701.

\bibitem{kim2000computation}
Y.-T. Kim, N.~Goldenfeld, J.~Dantzig, Computation of dendritic microstructures
  using a level set method, Physical Review E 62~(2) (2000) 2471.

\bibitem{gibou2003level}
F.~Gibou, R.~Fedkiw, R.~Caflisch, S.~Osher, A level set approach for the
  numerical simulation of dendritic growth, Journal of Scientific Computing
  19~(1) (2003) 183--199.

\bibitem{steinbach1999three}
I.~Steinbach, C.~Beckermann, B.~Kauerauf, Q.~Li, J.~Guo, Three-dimensional
  modeling of equiaxed dendritic growth on a mesoscopic scale, Acta Materialia
  47~(3) (1999) 971--982.

\bibitem{souhar2016three}
Y.~Souhar, V.~F. De~Felice, C.~Beckermann, H.~Combeau, M.~Zalo{\v{z}}nik,
  Three-dimensional mesoscopic modeling of equiaxed dendritic solidification of
  a binary alloy, Computational Materials Science 112 (2016) 304--317.

\bibitem{viardin2017mesoscopic}
A.~Viardin, M.~Zalo{\v{z}}nik, Y.~Souhar, M.~Apel, H.~Combeau, Mesoscopic
  modeling of spacing and grain selection in columnar dendritic solidification:
  Envelope versus phase-field model, Acta Materialia 122 (2017) 386--399.

\bibitem{tourret2016three}
D.~Tourret, A.~Karma, Three-dimensional dendritic needle network model for
  alloy solidification, Acta Materialia 120 (2016) 240--254.

\bibitem{isensee2022convective}
T.~Isensee, D.~Tourret, Convective effects on columnar dendritic
  solidification--a multiscale dendritic needle network study, Acta Materialia
  (2022) 118035.

\bibitem{rappaz1993probabilistic}
M.~Rappaz, C.-A. Gandin, Probabilistic modelling of microstructure formation in
  solidification processes, Acta metallurgica et materialia 41~(2) (1993)
  345--360.

\bibitem{gandin1994coupled}
C.-A. Gandin, M.~Rappaz, A coupled finite element-cellular automaton model for
  the prediction of dendritic grain structures in solidification processes,
  Acta metallurgica et materialia 42~(7) (1994) 2233--2246.

\bibitem{boettinger2002phase}
W.~J. Boettinger, J.~A. Warren, C.~Beckermann, A.~Karma, Phase-field simulation
  of solidification, Annual review of materials research 32~(1) (2002)
  163--194.

\bibitem{tourret2022phase}
D.~Tourret, H.~Liu, J.~LLorca, Phase-field modeling of microstructure
  evolution: Recent applications, perspectives and challenges, Progress in
  Materials Science 123 (2022) 100810.

\bibitem{steinbach2009phase}
I.~Steinbach, Phase-field models in materials science, Modelling and simulation
  in materials science and engineering 17~(7) (2009) 073001.

\bibitem{karma1998quantitative}
A.~Karma, W.~Rappel, Quantitative phase-field modeling of dendritic growth in
  two and three dimensions, Physical Review E 57~(4) (1998) 4323.

\bibitem{echebarria2004quantitative}
B.~Echebarria, R.~Folch, A.~Karma, M.~Plapp, Quantitative phase-field model of
  alloy solidification, Physical review E 70~(6) (2004) 061604.

\bibitem{shibuta2015solidification}
Y.~Shibuta, M.~Ohno, T.~Takaki, Solidification in a supercomputer: from crystal
  nuclei to dendrite assemblages, Jom 67~(8) (2015) 1793--1804.

\bibitem{shimokawabe2011peta}
T.~Shimokawabe, T.~Aoki, T.~Takaki, T.~Endo, A.~Yamanaka, N.~Maruyama,
  A.~Nukada, S.~Matsuoka, Peta-scale phase-field simulation for dendritic
  solidification on the tsubame 2.0 supercomputer, in: Proceedings of 2011
  International Conference for High Performance Computing, Networking, Storage
  and Analysis, 2011, pp. 1--11.

\bibitem{elahi2022multiscale}
S.~Elahi, R.~Tavakoli, A.~Boukellal, T.~Isensee, I.~Romero, D.~Tourret,
  Multiscale simulation of powder-bed fusion processing of metallic alloys,
  Computational Materials Science 209 (2022) 111383.

\bibitem{chen2016three}
S.~Chen, G.~Guillemot, C.-A. Gandin, Three-dimensional cellular
  automaton-finite element modeling of solidification grain structures for
  arc-welding processes, Acta materialia 115 (2016) 448--467.

\bibitem{rai2016coupled}
A.~Rai, M.~Markl, C.~K{\"o}rner, A coupled cellular automaton--lattice
  boltzmann model for grain structure simulation during additive manufacturing,
  Computational Materials Science 124 (2016) 37--48.

\bibitem{koepf2019numerical}
J.~Koepf, D.~Soldner, M.~Ramsperger, J.~Mergheim, M.~Markl, C.~K{\"o}rner,
  Numerical microstructure prediction by a coupled finite element cellular
  automaton model for selective electron beam melting, Computational Materials
  Science 162 (2019) 148--155.

\bibitem{lian2019cellular}
Y.~Lian, Z.~Gan, C.~Yu, D.~Kats, W.~K. Liu, G.~J. Wagner, A cellular automaton
  finite volume method for microstructure evolution during additive
  manufacturing, Materials \& Design 169 (2019) 107672.

\bibitem{mohebbi2020implementation}
M.~S. Mohebbi, V.~Ploshikhin, Implementation of nucleation in cellular
  automaton simulation of microstructural evolution during additive
  manufacturing of al alloys, Additive Manufacturing 36 (2020) 101726.

\bibitem{teferra2021optimizing}
K.~Teferra, D.~J. Rowenhorst, Optimizing the cellular automata finite element
  model for additive manufacturing to simulate large microstructures, Acta
  Materialia 213 (2021) 116930.

\bibitem{ivantsov1947temperature}
G.~IVANTSOV, Temperature field around a spherical cylindrical and a cicular
  crystal growing in a supercooled melt, in: Dokl. Akad. Nauk SSSR, Vol.~58,
  1947, pp. 567--569.

\bibitem{kurz1986theory}
W.~Kurz, B.~Giovanola, R.~Trivedi, Theory of microstructural development during
  rapid solidification, Acta metallurgica 34~(5) (1986) 823--830.

\bibitem{carozzani20113d}
T.~Carozzani, H.~Digonnet, C.-A. Gandin, 3d cafe modeling of grain structures:
  application to primary dendritic and secondary eutectic solidification,
  Modelling and Simulation in Materials Science and Engineering 20~(1) (2011)
  015010.

\bibitem{wang2003model}
W.~Wang, P.~D. Lee, M.~Mclean, A model of solidification microstructures in
  nickel-based superalloys: predicting primary dendrite spacing selection, Acta
  materialia 51~(10) (2003) 2971--2987.

\bibitem{yin2011simulation}
H.~Yin, S.~Felicelli, L.~Wang, Simulation of a dendritic microstructure with
  the lattice boltzmann and cellular automaton methods, Acta Materialia 59~(8)
  (2011) 3124--3136.

\bibitem{choudhury2012comparison}
A.~Choudhury, K.~Reuther, E.~Wesner, A.~August, B.~Nestler, M.~Rettenmayr,
  Comparison of phase-field and cellular automaton models for dendritic
  solidification in al--cu alloy, Computational Materials Science 55 (2012)
  263--268.

\bibitem{eshraghi2015large}
M.~Eshraghi, B.~Jelinek, S.~D. Felicelli, Large-scale three-dimensional
  simulation of dendritic solidification using lattice boltzmann method, Jom
  67~(8) (2015) 1786--1792.

\bibitem{gandin1995grain}
C.~A. Gandin, M.~Rappaz, D.~West, B.~Adams, Grain texture evolution during the
  columnar growth of dendritic alloys, Metallurgical and Materials Transactions
  A 26~(6) (1995) 1543--1551.

\bibitem{rappaz1996prediction}
M.~Rappaz, C.~A. Gandin, J.-L. Desbiolles, P.~Thevoz, Prediction of grain
  structures in various solidification processes, Metallurgical and Materials
  Transactions A 27~(3) (1996) 695--705.

\bibitem{takatani2000ebsd}
H.~Takatani, C.-A. Gandin, M.~Rappaz, Ebsd characterisation and modelling of
  columnar dendritic grains growing in the presence of fluid flow, Acta
  materialia 48~(3) (2000) 675--688.

\bibitem{carter2000process}
P.~Carter, D.~Cox, C.-A. Gandin, R.~C. Reed, Process modelling of grain
  selection during the solidification of single crystal superalloy castings,
  Materials Science and Engineering: A 280~(2) (2000) 233--246.

\bibitem{gandin19973d}
C.-A. Gandin, M.~Rappaz, A 3d cellular automaton algorithm for the prediction
  of dendritic grain growth, Acta Materialia 45~(5) (1997) 2187--2195.

\bibitem{pineau2018growth}
A.~Pineau, G.~Guillemot, D.~Tourret, A.~Karma, C.-A. Gandin, Growth competition
  between columnar dendritic grains--cellular automaton versus phase field
  modeling, Acta Materialia 155 (2018) 286--301.

\bibitem{dorari2022growth}
E.~Dorari, K.~Ji, G.~Guillemot, C.-A. Gandin, A.~Karma, Growth competition
  between columnar dendritic grains--the role of microstructural length scales,
  Acta Materialia 223 (2022) 117395.

\bibitem{jadhav2019influence}
S.~D. Jadhav, S.~Dadbakhsh, L.~Goossens, J.~Kruth, J.~Van~Humbeeck,
  K.~Vanmeensel, Influence of selective laser melting process parameters on
  texture evolution in pure copper, Journal of Materials Processing Technology
  270 (2019) 47--58.

\bibitem{higashi2020selective}
M.~Higashi, T.~Ozaki, Selective laser melting of pure molybdenum: evolution of
  defect and crystallographic texture with process parameters, Materials \&
  Design 191 (2020) 108588.

\bibitem{glasner2001nonlinear}
K.~Glasner, Nonlinear preconditioning for diffuse interfaces, J Comput Phys
  174~(2) (2001) 695--711.

\bibitem{karayagiz2020finite}
K.~Karayagiz, L.~Johnson, R.~Seede, V.~Attari, B.~Zhang, X.~Huang, S.~Ghosh,
  T.~Duong, I.~Karaman, A.~Elwany, et~al., Finite interface dissipation phase
  field modeling of ni--nb under additive manufacturing conditions, Acta
  Materialia 185 (2020) 320--339.

\bibitem{gandin1999three}
C.~Gandin, J.-L. Desbiolles, M.~Rappaz, P.~Thevoz, et~al., A three-dimensional
  cellular automation-finite element model for the prediction of solidification
  grain structures, Metallurgical and Materials Transactions A 30~(12) (1999)
  3153--3165.

\bibitem{carozzani2012developpement}
T.~Carozzani, D{\'e}veloppement d'un mod{\`e}le 3d automate
  cellulaire-{\'e}l{\'e}ments finis (cafe) parall{\`e}le pour la pr{\'e}diction
  de structures de grains lors de la solidification d'alliages m{\'e}talliques,
  Ph.D. thesis, Ecole Nationale Sup{\'e}rieure des Mines de Paris (2012).

\bibitem{fleurisson2022hybrid}
R.~Fleurisson, O.~Senninger, G.~Guillemot, C.-A. Gandin, Hybrid cellular
  automaton-parabolic thick needle model for equiaxed dendritic solidification,
  Journal of Materials Science \& Technology 124 (2022) 26--40.

\bibitem{mullins1964stability}
W.~W. Mullins, R.~Sekerka, Stability of a planar interface during
  solidification of a dilute binary alloy, Journal of applied physics 35~(2)
  (1964) 444--451.

\bibitem{kurz1994rapid}
W.~Kurz, R.~Trivedi, Rapid solidification processing and microstructure
  formation, Materials Science and Engineering: A 179 (1994) 46--51.

\bibitem{carrard1992banded}
M.~Carrard, M.~Gremaud, M.~Zimmermann, W.~Kurz, About the banded structure in
  rapidly solidified dendritic and eutectic alloys, Acta metallurgica et
  materialia 40~(5) (1992) 983--996.

\bibitem{kurz1996banded}
W.~Kurz, R.~Trivedi, Banded solidification microstructures, Metallurgical and
  Materials Transactions A 27~(3) (1996) 625--634.

\bibitem{mckeown2016time}
J.~T. McKeown, K.~Zweiacker, C.~Liu, D.~R. Coughlin, A.~J. Clarke, J.~K.
  Baldwin, J.~W. Gibbs, J.~D. Roehling, S.~D. Imhoff, P.~J. Gibbs, et~al.,
  Time-resolved in situ measurements during rapid alloy solidification:
  Experimental insight for additive manufacturing, Jom 68~(3) (2016) 985--999.

\bibitem{pilgar2022effect}
C.~M. Pilgar, A.~M. Fernandez, S.~Lucarini, J.~Segurado, Effect of printing
  direction and thickness on the mechanical behavior of slm fabricated
  hastelloy-x, International Journal of Plasticity 153 (2022) 103250.

\end{thebibliography}

\newpage


\section*{Supplementary material}

# Grain growth competition during melt pool solidification – Comparing phase-field and cellular automaton models

## Supplementary Material

S.M. Elahi[a,b], R. Tavakoli[b], I. Romero[a,b], D. Tourret[b]

[a] *Universidad Politécnica de Madrid, E.T.S. de Ingenieros Industriales, Madrid, Spain.*

[b] *IMDEA Materials Institute, C/ Eric Kandel 2, 28906, Getafe, Madrid, Spain.*

**Contents**

## A. Analytical approximation of the temperature field in the melt pool region

Based on the finite elements (FE) simulations of laser melting of Inconel 718 alloy performed in Ref. [1], we approximate the temperature field in the melt pool region along longitudinal and cross-section using analytical functions. These functions represent the location of liquidus ($T_L$) and solidus ($T_S$) isotherms as ellipsis. The function used along the longitudinal section is presented and discussed in details in Ref. [1]. The function used along the cross section (schematized in Fig. A.1) is given by:

$$T(r,\theta) = T_0(\theta, t) + \left(T_S - T_0(\theta, t)\right) \frac{r}{r_S(\theta, t)} \tag{A.1}$$

$$r_S(\theta) = \sqrt{\frac{(w_S(t) d_S(t))^2}{(d_S(t)\cos(\theta))^2 + (w_S(t)\sin(\theta))^2}} \tag{A.2}$$

with

$$r = \sqrt{(x-x_0)^2 + (y-y_0)^2} \tag{A.3}$$

$$\theta = \tan^{-1}\left|\frac{y-y_0}{x-x_0}\right| \tag{A.4}$$

and

$$w_S(t) = w_0 \times \left[1 - (t/t_f)^{n_w}\right] \tag{A.5}$$

$$d_S(t) = d_0 \times \left[1 - (t/t_f)^{n_d}\right] \tag{A.6}$$

$$T_0(\theta, t) = \frac{T_1(\theta) + T_2(t)}{2} - \frac{T_1(\theta) - T_2(t)}{2} \times \tanh\left(\frac{t - \tau(\theta)}{\sigma(\theta)}\right) \tag{A.7}$$

$$T_1(\theta) = T_w - (T_w - T_d) \times \sin(\theta) \tag{A.8}$$

$$T_2(t) = T_L - R \times (t - t_L) \tag{A.9}$$

$$\tau(\theta) = \tau_w - (\tau_w - \tau_d) \times \sin(\theta) \tag{A.10}$$

$$\sigma(\theta) = \sigma_w - (\sigma_w - \sigma_d) \times \sin(\theta) \tag{A.11}$$

$$\tag{A.12}$$

Here, $w_S(t)$ and $d_S(t)$ represent, respectively, the half-width and depth of the solidus temperature ($T_S$) location, while $T_0(\theta, t)$ extrapolates the temperature at the center of the melt pool, i.e. at $(x, y) = (0, 0)$. The melt pool center temperature $T_0(\theta, t)$ transitions from an early stage $T_1(\theta)$ to a later stage $T_2(t)$ using a sigmoid function (Eq. (A.7)). $T_1(\theta)$ is extrapolated from liquidus and solidus temperature locations along the width ($T_w$) and depth ($T_d$) of the melt pool. $T_2(t)$ is fitted to the late stage ($T_0 < T_L$) temperature evolution of the temperature. Function $T_1(\theta)$ and the sigmoid transition time and duration (parametrized using $\tau(\theta)$ and $\sigma(\theta)$) are interpolated between the melt pool width ($\theta = 0$) and depth ($\theta = \pi/2$) directions using sine functions (Eqs (A.8), (A.10) and (A.11)).

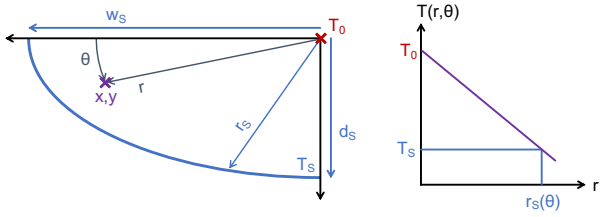

Figure A.1: Schematics of analytical temperature approximation $T(r,\theta)$: (left) location of elliptical solidus isotherm from the center $(x_0, y_0)$ of the melt pool, (right) radial interpolation between center temperature $T_0$ and solidus temperature $T_S$.

Hence, values of $T_w$, $T_d$, $\tau_w$, $\tau_d$, $\sigma_w$, and $\sigma_d$ are adjusted to fit the time evolution of the width and depth of the solidus ($T_S$) and liquidus ($T_L$) temperatures (Fig. A.2a), while $R$ is adjusted to match the cooling rate of $T_0$ once it falls below $T_L$ (Fig. A.2b), which can all be adjusted to match an arbitrary temperature profile, e.g. that calculated with FE in Ref. [1].

While arguably approximate, this analytical function has several advantages: (1) it allows a reasonable description of the time evolution of both solidus ($T_S$) and liquidus ($T_L$) width and depth, with an elliptical interpolation for $0 < \theta < \pi/2$ and a linear (radial) interpolation between $T_S$ and $T_L$ (where most of the solidification takes place); (2) it smoothly and continuously interpolates in time between $T_1$ and $T_2$ when $T_0 \approx T_L$, as long as appropriate values are chosen for $\tau_w$, $\tau_d$, $\sigma_w$, and $\sigma_d$ for this transition to occur early enough; and (3) it allows avoiding infinite temperature gradient at the melt pool center when $T_0$ gets close to $T_S$, possible with simpler function (e.g. of the form used in the longitudinal section), as long as the late-stage cooling rate $R$ is high enough.

Figures below illustrate the equations above adjusted to the published FE results using $t_L = 0.00195$ s, $t_f = 0.00255$ s, $w_0 = 128$ μm, $d_0 = 96$ μm, $n_w = 20$, $n_d = 3$, $T_w = 2700$ K, $T_d = 2450$ K, $\tau_w = 0.00115$ s, $\tau_d = 0.0007$ s, $\sigma_w = 0.00036$ s, $\sigma_d = 0.0005$ s, and $R = 1.2 \times 10^5$ K/s, considering $T_L = 1625$ K and $T_S = 1554$ K. While the peculiar shape of the solidus isotherm – namely the cusps due to the lower thermal conductivity of the powder bed on the sides compared to the high conductivity of the dense solidified metal at the bottom [1] – cannot be accurately modeled using elliptical isotherms, the figure shows that the time evolution of the solidus and liquidus isotherms is reasonably approximated by the chosen function.

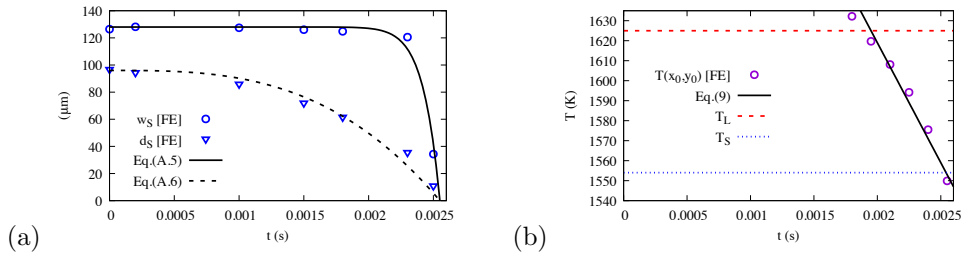

Figure A.2: (a) Time evolution of solidus isotherm half-width ($w_S$) and depth $d_S$ comparing FE results (symbols) and Eqs. (A.5)-(A.5) (lines). (b) Time evolution of the temperature at the center for the melt pool $T_0$ between $T_L$ and $T_S$, comparing FE results (symbols) with Eq. (A.9) (solid line).

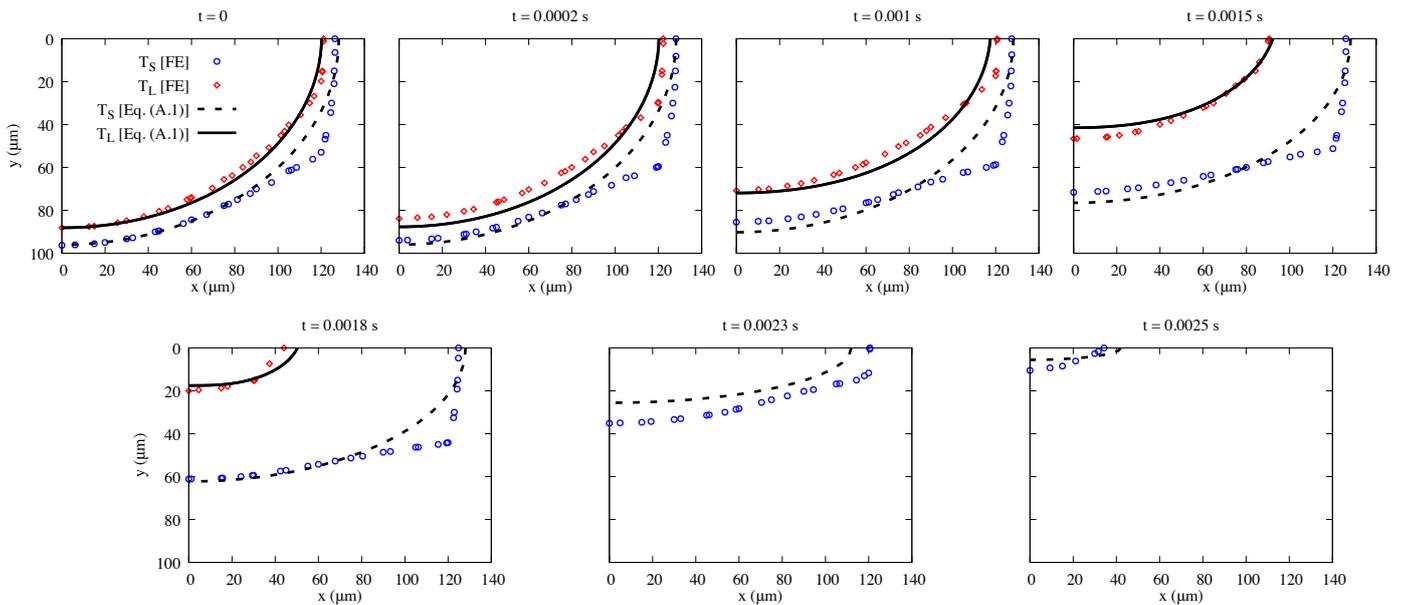

Figure A.3: Time evolution (snapshots) of the locations of the solidus and liquidus isotherms in a cross section, comparing FE results of Ref. [1] (symbols) to Eq. (A.1) (lines).

**B. Cross section grain maps**

See next pages.

**C. Computational performance**

Tables below summarize the computational performance of the various simulations in terms of simulation (wall) time and speed-up (wall time ratio) from cellular automaton (CA) compared to phase-field (PF) simulations. Note that we are comparing PF simulations performed on multiple GPUs (each with thousands of CUDA cores) to CA simulations performed on a single core on a desktop or laptop computer. For a similar simulation, the speed-up from a single core to a single-GPU parallel implementation is roughly of order $20\times$ and the strong scaling with the number of GPUs (here 2 to 8) was also observed to be nearly linear – both effects advantaging the PF performance by a significant amount in the tables below. Also note the significant differences in computing time when using different GPU models, e.g. with a factor of $\approx 1.4\times$ between the cross-section reference PF calculations, performed on two RTX3090 (faster) and the wide/shallow PF simulations performed on two RTX2080Ti (slower), while the simulation size is nearly identical.

| Method: | Phase-Field | Cellular Automaton | | |
|---|---|---|---|---|
| Hardware: | $8 \times$ RTX2080Ti | $1 \times$ Intel Xeon 6130 core (2.10 GHz) | | |
| | Wall time | $h_{\text{CA}}/h_{\text{PF}}$ | Wall time | Speed-up PF/CA |
| | 9d16h | 6 | 10d10h | 0.93 |
| | (232h) | 12 | 1d08h | 7.2 |
| | | 24 | 4h02m | 58 |
| | | 48 | 32m | 430 |
| | | 96 | 4m | 3 500 |
| | | 192 | 33s | 25 000 |
| | | 384 | 4s | 210 000 |
| | | 768 | 1s | 840 000 |

Table C.1: Computational performance for **longitudinal section** simulations.

| Method: | Phase-Field | Cellular Automaton | | |
|---|---|---|---|---|
| Hardware: | $2 \times$ RTX3090 | $1 \times$ Intel Xeon 6130 core (2.10 GHz) | | |
| | Wall time | $h_{\text{CA}}/h_{\text{PF}}$ | Wall time | Speed-up PF/CA |
| | 1d19h | 3 | 4d05h | 0.43 |
| | (43h) | 6 | 13h02m | 3.3 |
| | | 9 | 4h20m | 9.9 |
| | | 12 | 1h35m | 27 |
| | | 18 | 33m | 78 |
| | | 24 | 11m | 230 |
| | | 36 | 3m | 860 |
| | | 48 | 81s | 1 900 |
| | | 72 | 29s | 5 300 |
| | | 96 | 10s | 15 000 |
| | | 114 | 3s | 52 000 |

Table C.2: Computational performance for **cross-section** simulations for **reference melt pool** size.

3

| Method: | Phase-Field | Cellular Automaton | | |
|---|---|---|---|---|
| Hardware: | 2 × RTX2080Ti | 1 × Intel Xeon 6130 core (2.10 GHz) | | |
| | Wall time | $h_{\rm CA}/h_{\rm PF}$ | Wall time | Speed-up PF/CA |
| | 2d11h | 3 | 7d21h | 0.31 |
| | (59h) | 6 | 1d00h | 2.4 |
| | | 9 | 8h10m | 7.2 |
| | | 12 | 3h25m | 19 |
| | | 18 | 1h08m | 52 |
| | | 24 | 26m | 140 |
| | | 36 | 10m | 350 |
| | | 48 | 3m | 1 200 |
| | | 72 | 58s | 3 700 |
| | | 96 | 19s | 11 000 |
| | | 114 | 6s | 35 000 |

Table C.3: Computational performance for **cross-section** simulations for **wide & shallow melt pool** size.

| Method: | Phase-Field | Cellular Automaton | | |
|---|---|---|---|---|
| Hardware: | 2 × RTX3090 | 1 × Intel Xeon 6130 core (2.10 GHz) | | |
| | Wall time | $h_{\rm CA}/h_{\rm PF}$ | Wall time | Speed-up PF/CA |
| | 1d18h | 3 | 2d05h | 0.79 |
| | (42h) | 6 | 6h17m | 6.7 |
| | | 9 | 2h12m | 19 |
| | | 12 | 45m | 56 |
| | | 18 | 13m | 190 |
| | | 24 | 5m | 500 |
| | | 36 | 93s | 1 600 |
| | | 48 | 40s | 3 800 |
| | | 72 | 14s | 11 000 |
| | | 96 | 5s | 30 000 |
| | | 114 | 2s | 76 000 |

Table C.4: Computational performance for **cross-section** simulations for **deep & narrow melt pool** size.

## References

[1] S.M. Elahi, R. Tavakoli, A.K. Boukellal, T. Isensee, I. Romero, D. Tourret, *Multiscale simulation of powder-bed fusion processing of metallic alloys*, Computational Materials Science 209, 111383 (2022).

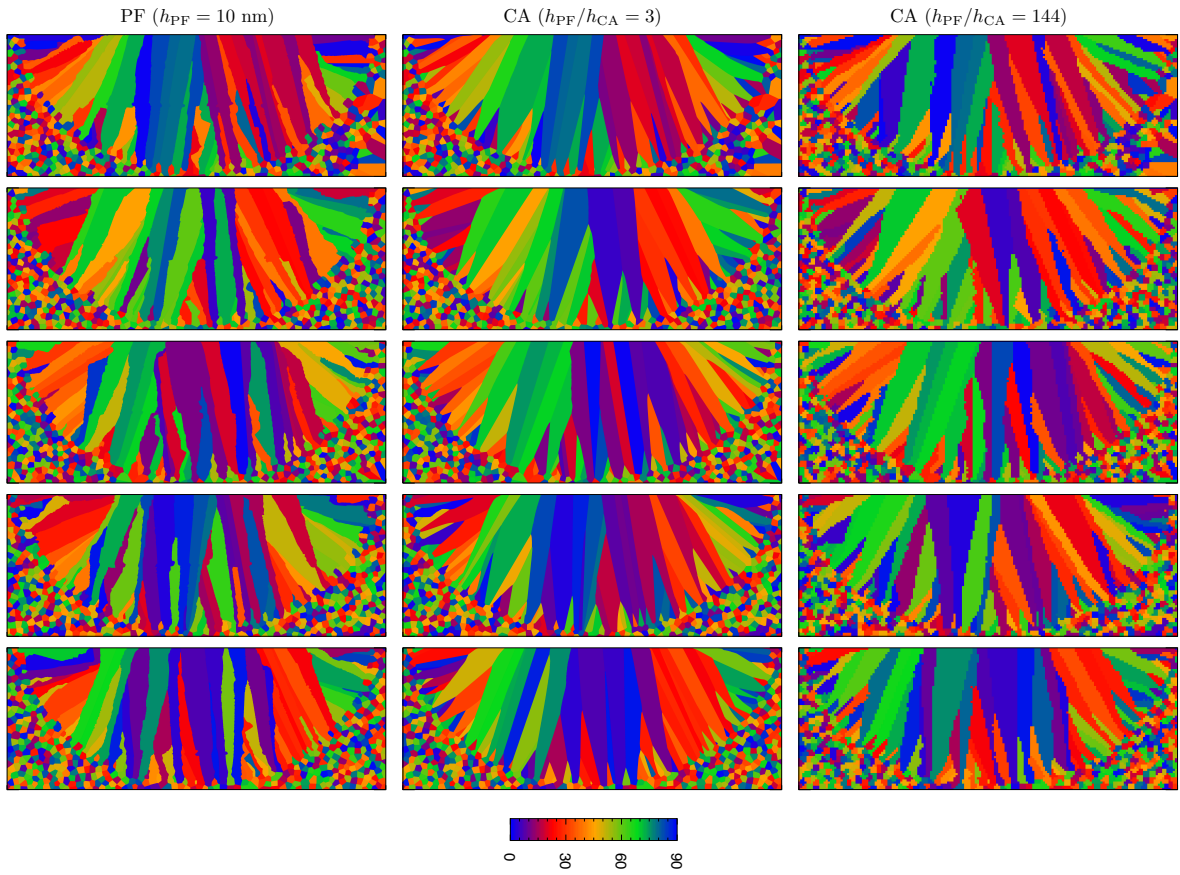

Figure C.1: Simulations of cross-section melt pool solidification for the reference case (i.e. thermal field as extracted from FE results) for five different initial grain distributions (rows) as predicted by phase-field (left column), and cellular automaton with fine (central column) and coarse (right column) spatial discretization. Domain size $\approx 257\,\mu\text{m} \times 97\,\mu\text{m}$.

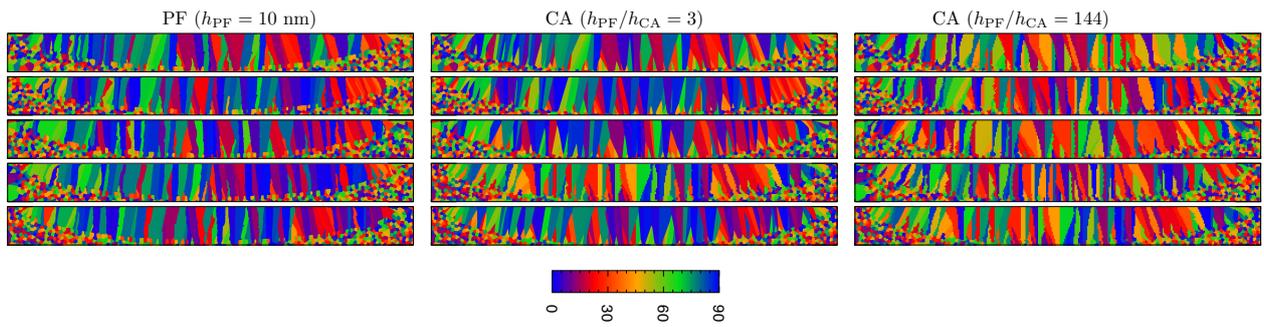

Figure C.2: Simulations of cross-section melt pool solidification for the reference case (i.e. thermal field as extracted from FE results) for five different initial grain distributions (rows) as predicted by phase-field (left column), and cellular automaton with fine (central column) and coarse (right column) spatial discretization. Domain size $\approx 513\,\mu$m $\times$ 49 $\mu$m.

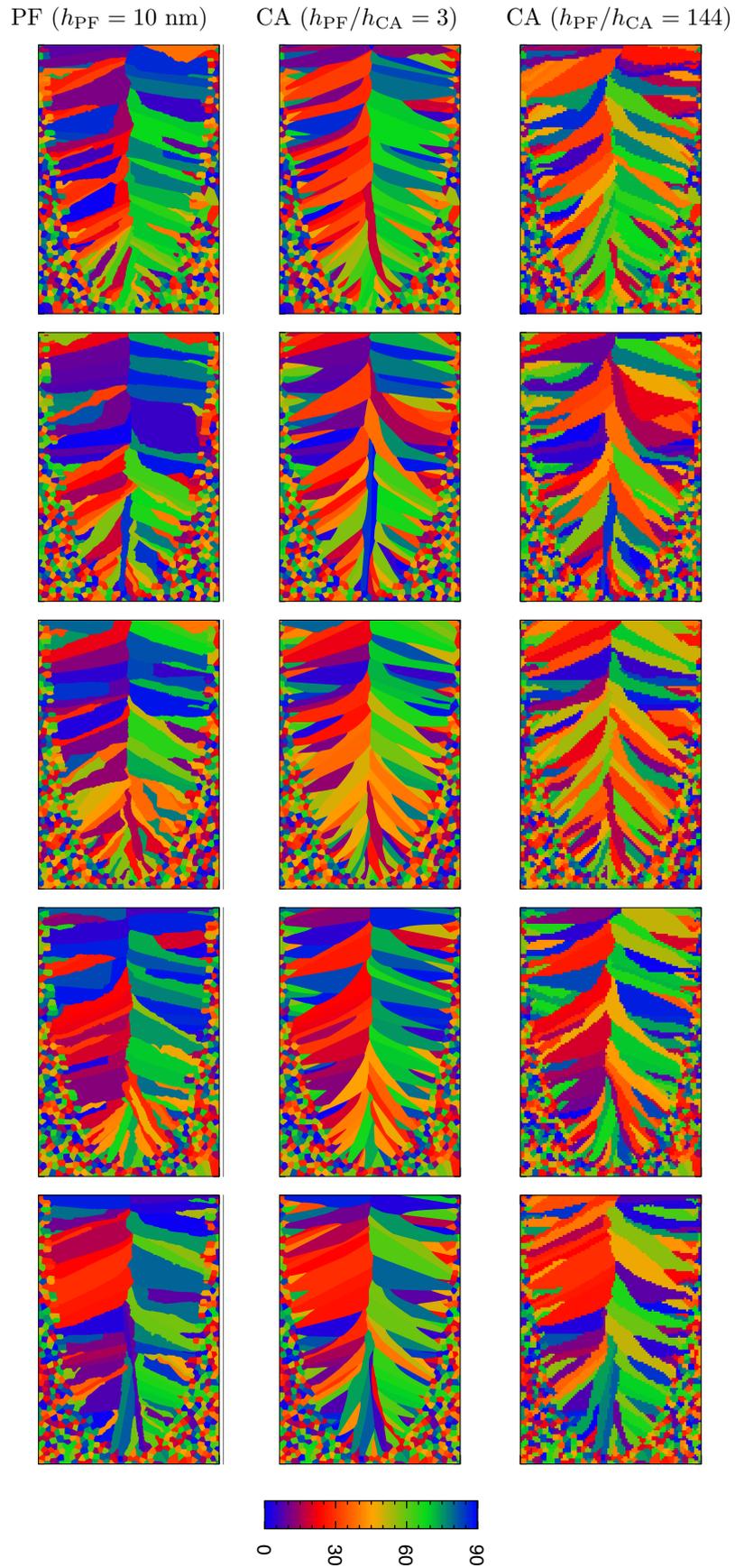

Figure C.3: Simulations of cross-section melt pool solidification of the deeper/narrower melt pool for five different initial grain distributions (rows) as predicted by phase-field (left column), and cellular automaton with fine (central column) and coarse (right column) spatial discretization. Domain size ≈ 129 $\mu$m × 193 $\mu$m.



\end{document}